\documentclass[letterpaper,floatfix,aps,prb,amsmath,twocolumn,
showpacs, nofootinbib]{revtex4}

\usepackage{epsfig}
\usepackage{verbatim}
\usepackage{epsf}
\usepackage{array}
\usepackage{hyperref}

\newcommand{\be}{\begin{equation}}
\newcommand{\ee}{\end{equation}}
\newcommand{\bea}{\begin{eqnarray}}
\newcommand{\eea}{\end{eqnarray}}

\renewcommand{\Re}{\mathrm{Re}\,}
\renewcommand{\Im}{\mathrm{Im}\,}

\newcommand{\eref}[1]{Eq.~(\ref{#1})}
\newcommand{\esref}[1]{Eqs.~(\ref{#1})}
\newcommand{\rref}[1]{(\ref{#1})}

\newcommand{\ocite}[1]{Ref.~\onlinecite{#1}}

\newcommand{\qp}{\mathrm{qp}}

\newcommand{\EZO}{\omega_{10}}
\newcommand{\ka}{x_{\mathrm{qp}}}
\newcommand{\kaA}{x^{\mathrm{A}}_{\mathrm{qp}}}
\newcommand{\Gif}{\Gamma_{i\to f}}

\begin{document}

\title{Relaxation and frequency shifts induced by quasiparticles in superconducting qubits}

\author{G. Catelani}
\author{R. J. Schoelkopf}
\author{M. H. Devoret}
\author{L. I. Glazman}
\affiliation{Departments of Physics and Applied Physics, Yale
University, New Haven, CT 06520, USA}

\begin{abstract}
As low-loss non-linear elements, Josephson junctions are the building blocks of superconducting qubits.
The interaction of the qubit degree of freedom with the quasiparticles tunneling through
the junction represent an intrinsic relaxation mechanism. We develop a general theory for the qubit
decay rate induced by quasiparticles, and we study its dependence on the magnetic flux used to
tune the qubit properties in devices such as the phase and flux qubits, the split transmon, and the fluxonium.
Our estimates for the decay rate apply to both thermal equilibrium and non-equilibrium quasiparticles.
We propose measuring the rate in a split transmon to obtain information on the possible non-equilibrium
quasiparticle distribution. We also derive expressions for the shift in qubit frequency in the presence of
quasiparticles.
\end{abstract}

\date{\today}


\maketitle

\section{Introduction}
\label{sec:intro}

The operability of a quantum device as a qubit requires long coherence times in comparison to
the gate operation time.~\cite{divincenzo} Over the years, longer coherence times
in superconducting qubits have been achieved by designing new systems in
which the decoupling of the quantum oscillations of the order parameter from other low-energy
degrees of freedom is enhanced. For example, in a transmon qubit~\cite{transmon} the sensitivity
to background charge noise is suppressed relative to that of a Cooper pair box. Irrespective
of the particular design, in any superconducting device the qubit degree of freedom can exchange energy with
quasiparticles. This intrinsic relaxation mechanism is suppressed in thermal equilibrium at temperatures
much lower than the critical temperature, due to the exponential depletion of the quasiparticle population.
However, both in superconducting qubits~\cite{Martinis} and
resonators~\cite{klapwijk} nonequilibrium quasiparticles have been observed which can lead to relaxation even
at low temperatures. In this paper we study the quasiparticle
relaxation mechanism in qubits based on Josephson junctions, both for equilibrium and nonequilibrium
quasiparticles.

Quasiparticle relaxation in a Cooper pair box was considered in \ocite{lutchyn1}.
In this system the charging energy is large compared to the Josephson energy and
quasiparticle poisoning~\cite{matveev,joyez} is the elementary
process of relaxation: a quasiparticle entering the
Cooper pair box changes the parity (even or odd) of the state, bringing the qubit
out of the computational space consisting of two charge states of the same parity. More recently the theory
of \ocite{lutchyn1} was extended to estimate the effect of quasiparticles in a
transmon.\cite{transmon} In this case the dominant energy scale is
the Josephson energy, so that quantum
fluctuations of the phase are relatively small, while
the uncertainty of charge in the qubit states is significant. As mentioned above, the
advantage of the transmon is its low sensitivity to charge
noise. The possible role of nonequilibrium quasiparticles  in superconducting qubits was
investigated in \ocite{Martinis}. While
the properties of many superconducting qubits -- the phase and flux
qubits,~\cite{Dev_rev} the split transmon,
and the newly developed fluxonium~\cite{flux_exp} --
can be tuned by an external
magnetic flux, the effect of the latter on the quasiparticle relaxation rate has not been
previously analyzed. Elucidating the role of flux is the main goal of this work.
In particular, we show that studying the flux dependence of the relaxation rate can
provide information on the presence of nonequilibrium quasiparticles.

The paper is organized as follows: in the next section we present results for the admittance
of a Josephson junction and the general approach to calculate the
decay rate and energy level shifts due to quasiparticles in a qubit
with a single Josephson junction.
In Sec.~\ref{sec:semi} we consider a weakly
anharmonic qubit, such as phase qubit or transmon,
and relate its decay rate, quality factor, and frequency shift to the admittance
of the junction. The cases of a Cooper pair box (large charging energy)
and of a flux qubit with large Josephson energy are examined in Sec.~\ref{sec:further}.
Some of the results presented in Secs.~\ref{sec:th_s}-\ref{sec:further} have been
reported previously\cite{prl} in a brief format.
In Sec.~\ref{sec:multi} we describe the generalization to multi-junction
systems and study, as concrete examples, the two-junction split transmon and the many-junction fluxonium.
We summarize the present work in Sec.~\ref{sec:summ}.
Throughout the paper, we use units $\hbar=k_B =1$ (except otherwise noted).

\section{General theory for a single-junction qubit}
\label{sec:th_s}

\begin{figure}[b]
\begin{center}
\includegraphics[width=0.39\textwidth]{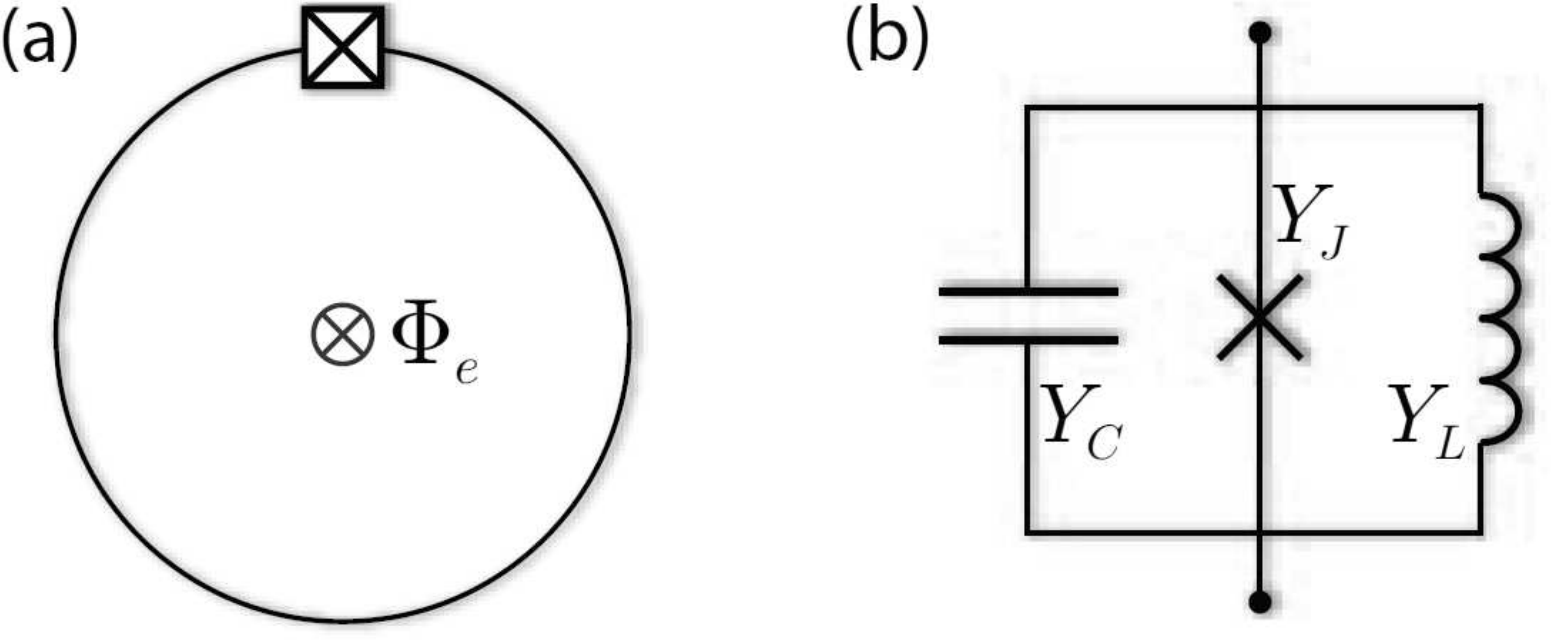}
\end{center}
\caption{(a) Schematic representation of a qubit controlled by a magnetic flux,
see \eref{Hphi}.
(b) Effective circuit diagram with three parallel elements
-- capacitor, Josephson junction, and inductor -- characterized by their respective admittances.}
\label{fig1}
\end{figure}

We consider a Josephson junction closed by an inductive loop, see Fig.\ref{fig1}.
The low-energy effective Hamiltonian of the system
can be separated into three parts
\be\label{Htot}
\hat{H} = \hat{H}_{\varphi} + \hat{H}_{\qp} + \hat{H}_{T} \, .
\ee
The first term determines the dynamics of the phase degree of freedom in the
absence of quasiparticles
\be\label{Hphi}
\hat{H}_{\varphi} = 4E_C \left(\hat{N}-n_g\right)^2 -E_J \cos \hat\varphi
+ \frac12 E_L\!\left(\hat\varphi-2\pi\Phi_e/\Phi_0\right)^2,
\ee
where $\hat{N}=-id/d\varphi$ is the number operator of Cooper
pairs passed across the junction, $n_g$ is the dimensionless gate voltage, $\Phi_e$ is the external flux
threading the loop, $\Phi_0=h/2e$ is the flux quantum, and the
parameters characterizing the qubit are the charging energy $E_C$,
the Josephson energy $E_J$, and the inductive energy $E_L$.

The second term in \eref{Htot} is the sum of the BCS Hamiltonians for quasiparticles in the leads
\be\label{Hqp}
\hat{H}_{\qp} = \sum_{j=L,R} \hat{H}_{\qp}^j \, , \quad \hat{H}_{\qp}^j = \sum_{n,\sigma} \epsilon_{n}^j
\hat\alpha^{j\dagger}_{n\sigma} \hat\alpha^j_{n\sigma},
\ee
where $\hat\alpha^j_{n\sigma}$($\hat\alpha^{j\dagger}_{n\sigma}$) are quasiparticle annihilation (creation)
operators and $\sigma=\uparrow, \downarrow$ accounts for spin. The
quasiparticle energies are $\epsilon^j_{n} =
\sqrt{(\xi_{n}^j)^2+(\Delta^j)^2}$, with $\xi_{n}^j$ and $\Delta^j$ being the
single-particle energy level $n$ in the normal state of lead $j$, and
the gap parameter in that lead, respectively.
The occupations of the quasiparticle states are described by the distribution functions
\be
f^j(\xi^j_{n})= \langle\!\langle \hat\alpha_{n\uparrow}^{j\dagger} \hat\alpha^j_{n\uparrow}\rangle\!\rangle_\qp
= \langle\!\langle \hat\alpha_{n\downarrow}^{j\dagger} \hat\alpha^j_{n\downarrow}\rangle\!\rangle_\qp\, ,
\quad j=L,R\, ,
\ee
assumed to be independent of spin; double angular brackets
$\langle\!\langle \ldots \rangle\!\rangle_\qp$ denote averaging over the
quasiparticle states. Hereinafter we assume for simplicity equal gaps in the leads,
$\Delta^L = \Delta^R \equiv \Delta$.

The last term in \eref{Htot} describes quasiparticle tunneling across the junction
and couples the phase and quasiparticle degrees of freedom
\bea\label{HT}
\hat{H}_{T} = \tilde{t}\!\!\sum_{n,m,\sigma}\!\!
\left(e^{i\frac{\hat\varphi}{2}} u_{n}^L u_{m}^R
- e^{-i\frac{\hat\varphi}{2}} v_{m}^R v_{n}^L
\right)\hat\alpha_{n\sigma}^{L\dagger} \hat\alpha^R_{m\sigma}
+ \text{H.c.}\quad
\eea
The electron tunneling amplitude $t$ in this equation determines the junction
conductance, $g_T = 4\pi e^2 \nu^L \nu^R \tilde{t}^2$
in the tunneling limit $\tilde{t}\ll 1$ which we are considering. From now on, we assume identical
densities of states per spin direction in the leads,
$\nu^L\!=\!\nu^R\!=\!\nu_0$.
The Bogoliubov amplitudes $u^j_{n}$, $v^j_{n}$
can be taken real, since \eref{HT} already accounts explicitly for
the phases of the order parameters in the leads
via the gauge-invariant phase difference~\cite{BP} in the exponentials.
Accounting for the Josephson effect and quasiparticles dynamics by
Eqs.~(\ref{Hphi})-(\ref{HT}) is possible as long as
the qubit energy $\omega$ and characteristic energy $\delta E$ of quasiparticles
(as determined by their distribution function and measured from $\Delta$) are small compared
to $\Delta$:~\cite{lutchyn1} $\omega, \delta E \ll 2\Delta$. In this
low-energy limit, we may further approximate $u^j_{m} \simeq v^j_{n} \simeq
1/\sqrt{2}$.
Then the operators $e^{\pm i\hat\varphi/2}$ in \eref{HT},
which describe transfer of charge $\pm e$ across the junction,
combine to give
\be\label{HTle}
\hat{H}_{T} = \tilde{t}\!\!\sum_{n,m,\sigma}\!\!
i\sin \frac{\hat\varphi}{2}
\hat\alpha_{n\sigma}^{L\dagger} \hat\alpha^R_{m\sigma}
+ \text{H.c.}
\ee
Starting from this low-energy tunneling
Hamiltonian, in the next section we calculate the dissipative part
of the junction admittance.

\subsection{Response to a classical time-dependent phase}

We consider here
the ``classical'' dissipative response of a Josephson junction
to a small ac bias to show that \eref{HTle} correctly accounts for the known~\cite{BP}
junction losses in the low-energy regime. These
``classical'' losses are directly related to the decay rate in the quantum regime,
as we explicitly show in the next section.

We assume a time-dependent bias
$v(t)= v \cos(\omega t)$ of
frequency $\omega>0$ superimposed to a fixed phase difference $\varphi_0$.
In other words, we take the phase to be a time-dependent number which, by the Josephson equation
$d\varphi/dt = 2e v(t)$, has the form
\be\label{ph_time}
\varphi(t) = \varphi_0 + \frac{2ev}{\omega} \sin (\omega t)\, .
\ee
Here we focus on the linear in $v$ response in the low-energy regime. Expressions for
the current through the junction valid beyond
linear response can be found, for example, in \ocite{BP}.
Substituting \eref{ph_time} into
\eref{HTle}, expanding for small $v$,
and keeping the linear term, we find for the time-dependent perturbation
$\delta\hat{H}(t)$ causing the dissipation
\be\label{Hac}\begin{split}
\delta\hat{H}(t) & = \hat{H}_{AC} \sin (\omega t)\, , \\
\hat{H}_{AC} & =
i \tilde{t}
\cos\frac{\varphi_0}{2} \frac{ev}{\omega}
\sum_{n,m,\sigma} \hat\alpha_{n\sigma}^{L\dagger} \hat\alpha^R_{m\sigma}
+ \text{H.c.}
\end{split}\ee
The average dissipated power can be calculated using Fermi's golden rule:
it is given by the product of the transition rate times
the energy change in a transition between quasiparticle states caused by the perturbation.
The energy change in a transition is $\pm\omega$ by energy conservation, with
the two signs corresponds to the events
giving energy to or taking energy from the system. The average power $P$ is
\bea\label{P0}
&& P  =  2\pi\sum_{\left\{\lambda\right\}_\qp} \langle\!\langle \left|\langle \left\{\lambda\right\}_\qp
 |\hat{H}_{AC}|\left\{\eta\right\}_\qp\rangle\right|^2 \omega \\
&& \times \left[ \delta (E_{\lambda,\qp} - E_{\eta,\qp} - \omega)
-  \delta (E_{\lambda,\qp} - E_{\eta,\qp} + \omega)\right]
\rangle\!\rangle_\qp \, ,\nonumber
\eea
where $E_{\eta,\qp}$ and $E_{\lambda,\qp}$ are the total energies of the
quasiparticles in their respective initial $\{\eta\}_\qp$ and final
$\{\lambda\}_\qp$ states. We use \eref{Hac} to evaluate the matrix element, average over initial quasiparticle
states, and sum over final states to find
\be
P = \frac{1}{2} \Re Y_J(\omega,\varphi_0) v^2
\ee
with\cite{cosphi}
\be\label{ReYJ}
\Re Y_J(\omega,\varphi) = \frac{1+\cos\varphi}{2} \Re Y_\qp (\omega).
\ee
Here $\Re Y_\qp$ is the real part of the quasiparticle contribution to the junction admittance
at zero phase difference,
\be\label{ReYqp}\begin{split}
\Re Y_\qp(\omega) = & g_T \frac{2\Delta}{\omega} \int_0^\infty\!dx \,
\frac{1}{\sqrt{x}\sqrt{x+\omega/\Delta}} \\ & \left[f_E\left((1+x)\Delta\right)
-f_E\left((1+x+\omega/\Delta)\Delta\right)\right].
\end{split}\ee
In deriving these formulas we have
approximated the standard BCS density of states functions as
\be\label{app_dos}
\frac{\epsilon}{\sqrt{\epsilon^2-\Delta^2}}, \,
\frac{\Delta}{\sqrt{\epsilon^2-\Delta^2}} \sim \sqrt{\frac{\Delta}{2(\epsilon-\Delta)}}
\equiv \frac{1}{\sqrt{2x}}
\ee
and taken equal quasiparticle occupations in the two leads,
$f^L = f^R\equiv f$; we use this simplifying assumption throughout the paper.
We indicate with $f_E$ the energy mode of the distribution function
\be\label{fE}
f_E(\epsilon) = \frac12 \left[f(\xi) + f(-\xi) \right],
\ee
where $\epsilon = \sqrt{\xi^2+\Delta^2}$.
Equation \rref{ReYJ} for the real part of the admittance, valid at $\omega>0$, agrees
with the linear response, low-energy limit of the non-linear $I$-$V$ characteristic
presented in \ocite{BP}. Extension to $\omega<0$ is found by noticing that
$\Re Y_\qp$ is an even function of frequency.

In thermal equilibrium and at low temperatures $T \ll \Delta$
the distribution function can be approximated as
\be\label{f_th}
f_E(\epsilon) \simeq e^{-\epsilon/T},
\ee
and \eref{ReYqp} gives, at arbitrary ratio $\omega/T$,
\be\label{ReYqp_te}\begin{split}
\Re Y_\qp^{eq} (\omega)  = g_T\frac{2\Delta}{\omega}
e^{-\Delta/T} e^{\omega/2T} K_0\left(\frac{|\omega|}{2T}\right)\left[1-e^{-\omega/T}\right].
\end{split}\ee
Here $K_0$ is the modified Bessel function of the second kind with asymptotes
\be\label{K0_as}
K_0 (x) \simeq \left\{\begin{array}{lll}
e^{-x}\sqrt{\pi/2x}\,, \quad & & x \gg 1 \\
\ln 2/x - \gamma_E\, , & & x \ll 1
\end{array}\right.
\ee
with $\gamma_E $ the Euler gamma.

For a generic distribution function, we can relate $\Re Y_\qp$ to the density of quasiparticle $n_\qp$
in the high-frequency regime
$\omega \gg \delta E$, where $\delta E$ indicates the characteristic energy of quasiparticle (measured
from the gap) above which the occupation of the quasiparticle states can be neglected; in thermal
equilibrium $\delta E \sim T$. Under the assumption $\omega \gg \delta E$ we obtain from \eref{ReYqp}
\be\label{ReYqp_hf}
\Re Y_\qp^{hf} (\omega) = \frac12 \ka \, g_T \left(\frac{2\Delta}{|\omega|}\right)^{3/2},
\ee
where
\be\label{xqp_def}
x_\qp = \frac{n_\qp}{2\nu_0 \Delta}
\ee
is the quasiparticle density normalized to the Cooper pair density and
\be\label{nqp_def}
n_\qp = 2\sqrt{2}\nu_0\Delta \int_0^\infty\frac{dx}{\sqrt{x}}
f_E((1+x)\Delta)
\ee
is the density written using the approximation in \eref{app_dos}. Note that in thermal equilibrium
at low temperatures, \eref{f_th}, we have
\be\label{nqp_th}
n_\qp^{eq} = 2\nu_0 \sqrt{2\pi\Delta T} e^{-\Delta/T}.
\ee
Then using \eref{K0_as}, it is easy to check that
for $T\ll \omega$ \eref{ReYqp_te} takes the form given in \eref{ReYqp_hf}.

The real and imaginary parts of the admittance satisfy the Kramers-Kr\"onig relations. However, when taking
the Kramers-Kr\"onig transform of the real part, a purely inductive contribution to the imaginary part can
be missed. Indeed, at low energies the complex junction admittance (obtained from the expressions
in~\ocite{BP}) can be written as
\be\label{YJ}
Y_J(\omega,\varphi) = \frac{1-2\kaA}{i\omega L_J} \cos\varphi + Y_\qp(\omega) \frac{1+\cos\varphi}{2}\, ,
\ee
where
\be\label{alp_def}
\kaA = f_E(\Delta)
\ee
can be interpreted as the population of the Andreev bound states~\cite{beenakker} and
the inverse of the Josephson inductance is
\be\label{LJ_def}
\frac{1}{L_J} = g_T \pi \Delta_\qp
\ee
(the subscript $\qp$ in $\Delta_\qp$ is used to indicate that in this expression
it may be necessary to account for the effect of quasiparticles on the gap, see Secs.~\ref{sec:encorr}
and \ref{sec:freq_sh}).

Unlike the Andreev states, free quasiparticles contribute to both dissipative and non-dissipative parts of
the total admittance $Y_J$ via the complex term $Y_\qp$.
The real part of the quasiparticle
admittance is defined in \eref{ReYqp}, while its imaginary part is given by
the Kramers-Kr\"onig transform of that expression,
\be\begin{split}
\Im Y_\qp(\omega) =  -g_T \frac{2\Delta}{\omega} \frac{P}{\pi}
\int_0^\infty\!\frac{dx}{\sqrt{x}}\int_0^\infty\!\frac{dy}{\sqrt{y}}
\Big[f_E\left((1+x)\Delta\right)
\\  -f_E\left((1+y)\Delta\right)\Big]
\left[\frac{1}{x-y+\omega/\Delta} - \frac{1}{x-y}\right],
\end{split}\ee
where $P$ denotes the principal part and $\omega > 0$. Using that $\Im Y_\qp$ is an odd
function of frequency, we can simplify the above expression to a form with a single
rather than double integral
\be
\Im Y_\qp(\omega) = g_T \frac{2\Delta}{\omega}\left[
\int_0^{|\omega|/\Delta}\!\!dx \,
\frac{f_E\left((1+x)\Delta\right)}{\sqrt{x}\sqrt{|\omega|/\Delta - x}}
-\pi \kaA
\right].
\ee

As discussed above for the real part, an analytic expression for $\Im Y_\qp$
can be obtained in thermal equilibrium,
\be\label{Y2_th}
\Im Y_\qp^{eq} (\omega) = -g_T \frac{2\Delta}{\omega}
e^{-\Delta/T}\pi\left[1 - e^{-|\omega|/2T} I_0 \left(\frac{|\omega|}{2T}\right)\right].
\ee
Here $I_0$ is the modified Bessel function of the first kind with asymptotes
\be\label{I0_as}
I_0(x) \simeq \left\{
\begin{array}{lll}
e^{x}\sqrt{1/2\pi x}\, , \quad & & x \gg 1 \\
1+ x^2/4\, , & & x \ll 1
\end{array}
\right. .
\ee

For arbitrary distribution
function satisfying the high-frequency condition $\omega \gg \delta E$ we find
\be\label{Y2_hf}
\Im Y_\qp^{hf}(\omega) = \frac12
g_T \, \frac{2\Delta}{\omega}
\left[  \ka \sqrt{\frac{2\Delta}{|\omega|}} - 2\pi\kaA \right].
\ee
Using \eref{nqp_th} and the large-$x$ limit in \eref{I0_as}, it is easy to show that
for $T\ll \omega$ \eref{Y2_th} reduces to the general expression in \eref{Y2_hf}.
In the high-frequency regime,
real and imaginary parts of the quasiparticle admittance can be combined into the complex admittance
\be\label{Yqp_hf}
Y_\qp^{hf} (\omega) = -\frac{2}{i\omega L_J}\left[\frac{\ka}{\pi}\sqrt{\frac{\Delta}{i\omega}}
- \kaA \right].
\ee
By substituting \eref{Yqp_hf} into \eref{YJ} we find that in the total admittance $Y_J$
the coefficient multiplying $\kaA$ is proportional to $(1-\cos\varphi)$ and vanishes for $\varphi=0$.
This is in agreement with the absence of Andreev bound states when there is no phase difference
across the junction.

\subsection{Transition rates}

The effects of the interaction between quasiparticles and qubit degree of freedom, \eref{HT}, can be treated
perturbatively in the tunneling amplitude $\tilde{t}$. The interaction makes possible,
for example, a transition between two qubit states (initial,
$|i\rangle$, and final, $|f\rangle$, differing in energy by amount $\omega_{if}>0$)
by exciting a quasiparticle during a tunneling event. The rate for the transition
between qubit states
can be calculated using Fermi's golden rule
\be\label{FGR}\begin{split}
\Gif = 2\pi \sum_{\{\lambda\}_\qp} &
\langle\!\langle  \left|\langle f , \{\lambda\}_\qp | \hat{H}_T |i, \{\eta\}_\qp \rangle\right|^2
 \\ & \ \times
 \delta\left(E_{\lambda,\qp}-E_{\eta,\qp}-\omega_{if}\right) \rangle\!\rangle_\qp \,.
\end{split}\ee
We remind that in our notation $E_{\eta,\qp}$ ($E_{\lambda,\qp}$) is the total energy of the
quasiparticles in their initial (final) state $\{\eta\}_\qp$ ($\{\lambda\}_\qp$), and
double angular brackets
$\langle\!\langle \ldots \rangle\!\rangle_\qp$ denote averaging over the
initial quasiparticle states whose occupation is determined by the distribution function.

In the low-energy regime we are considering, the transition rate factorizes into terms accounting separately
for qubit dynamic and quasiparticle kinetics
\be\label{wif_gen}
\Gif = \left|\langle f|\sin \frac{\hat\varphi}{2}|i\rangle\right|^2
S_\qp\left(\omega_{if}\right).
\ee
Equation \rref{wif_gen} is one of the main results of this work: it shows that
the qubit properties affect the transition rate via the wavefunctions $|i\rangle$,
and $|f\rangle$ entering the matrix element, while the quasiparticle kinetics is
accounted for by the quasiparticle current spectral density
$S_\qp$
\be\label{tF_def}\begin{split}
  S_\qp\left(\omega\right) = &
  \frac{16E_J}{\pi}
  \int_0^{\infty}\!\!\!dx\,
  \frac{1}{\sqrt{x}\sqrt{x+ \omega/\Delta}}\Big[ f_E \left((1+x)\Delta\right) \\
  & \times \left(1-f_E\left((1+x)\Delta+\omega\right)\right)
  \Big],
\end{split}\ee
where $\omega > 0$ and we used the relation
\be\label{EJdef}
E_J = g_T\Delta/8g_K
\ee
with $g_K = e^2/2\pi$ the conductance quantum.
The expression for $S_\qp$ at $\omega < 0$ is obtained by the replacements
$x \to x - \omega/\Delta$, $\omega \to -\omega$ in the integrand in \eref{tF_def}.

The spectral density $S_\qp$ depends on the detail of the distribution functions.
In thermal equilibrium
at low temperatures $T \ll \Delta$, using \eref{f_th}
we find
\be\label{Sqp_th}
S_\qp^{eq}(\omega) = \frac{16E_J}{\pi} e^{-\Delta/T} e^{\omega/2T} K_0\left(\frac{|\omega|}{2T}\right).
\ee
Note that the equality
\be
\frac{S_\qp^{eq}(-\omega)}{S_\qp^{eq}(\omega)} = e^{-\omega/T}
\ee
implies that in thermal equilibrium the transition rates are related by detailed balance,
\be
\frac{\Gamma_{f \to i}}{\Gif} = e^{-\omega_{if}/T}.
\ee

The similarity between \eref{Sqp_th} for $S_\qp$ and
\eref{ReYqp_te} for $\Re Y_\qp$ is not accidental.
In thermal equilibrium the following fluctuation-dissipation relation holds
\be
S_\qp^{eq}(\omega) + S_\qp^{eq}(-\omega) = \frac{\omega}{\pi} \frac{1}{g_K} \Re Y_\qp^{eq} (\omega)
\coth \left(\frac{\omega}{2T}\right).
\ee
Moreover, in the low-energy regime for an \textit{arbitrary} distribution function
the two quantities are also related by
\be\label{S_ReY}
S_\qp(\omega) - S_\qp(-\omega) = \frac{\omega}{\pi} \frac{1}{g_K} \Re Y_\qp (\omega)\, .
\ee

In the high-frequency regime $\omega \gg \delta E$, we can simplify the above relation to
\be\label{S_ReY_hf}
S_\qp^{hf}(\omega) = \frac{\omega}{\pi} \frac{1}{g_K} \Re Y_\qp^{hf} (\omega).
\ee
For the
transition rates this corresponds to neglecting the downward transitions with $\omega_{if} < 0$, in which
a quasiparticle looses energy to the qubit, compared to the upward ones. This is a good approximation since
the assumption $\omega \gg \delta E$ means that there are no quasiparticles with energy high enough to
excite the qubit. Equation \rref{S_ReY_hf} can be checked by comparing \eref{ReYqp_hf}
to
\be\label{Sqp_hf}
S_\qp^{hf}(\omega)  =  \ka
\frac{8E_J}{\pi}
\sqrt{\frac{2\Delta}{\omega}}
\ee
with $E_J$ given in \eref{EJdef} and the the normalized quasiparticle density $\ka$ in \eref{xqp_def}.

\subsection{Energy level corrections}
\label{sec:encorr}

In addition to causing transitions between qubit levels,
the quasiparticles affect the energy $E_i$ of each level $i$ of the system. We can distinguish two
quasiparticle mechanisms that modify the qubit spectrum and hence separate two terms in the correction
$\delta E_i$  to the energy,
\be\label{de_sep}
\delta E_i = \delta E_{i,E_J} + \delta E_{i,\qp} \, .
\ee

First, in the presence of quasiparticles the
Josephson energy takes the form
\be\label{EJqp}
E_{J,\qp} = \frac{g_T}{8 g_K}\Delta_\qp (1-2\kaA)
\ee
with $\kaA$ defined in \eref{alp_def}. As mentioned after \eref{LJ_def},
we use $\Delta_\qp$ to distinguish the self-consistent gap in
the presence of quasiparticles from the gap $\Delta$ when there are no quasiparticles.
At leading order in the quasiparticle density we have
\be\label{delqp}
\Delta_\qp \simeq \Delta \left(1-\ka\right).
\ee
Treating these modifications to the Josephson energy as perturbations, the
correction to the energy of level $i$ is
\be\label{de_EJ}
\delta E_{i,E_J} =  E_J \left(\ka + 2\kaA \right) \langle i |\cos \hat\varphi |i\rangle \, .
\ee

Second, the virtual transitions between the qubit
levels mediated by quasiparticle tunneling cause a correction
that can be expressed in terms of the matrix elements of $\sin\hat\varphi/2$ as
\be\label{de_qp}\begin{split}
\delta E_{i,\qp} = \sum_{k\neq i}
\left|\langle k | \sin\frac{\hat\varphi}{2}|i\rangle \right|^2
F_\qp \left(\omega_{ik}\right) \, ,
\end{split}\ee
where
\be
\omega_{ik} = E_k - E_i\, .
\ee
The derivation of the above formulas and the definition of function $F_\qp$ in terms of the
quasiparticle distribution function [\eref{Fqp_def}] are given in Appendix~\ref{app:encorr}.
Here we give the relation between $F_\qp$ and the imaginary part of the quasiparticle
impedance,
\be\label{F_ImY}
F_\qp(\omega) + F_\qp(-\omega)= -\frac{\omega}{2\pi}\frac{1}{g_K} \Im Y_\qp (\omega),
\ee
which we will use in the next section to obtain the quasiparticle-induced change in the qubit frequency.

\section{Single junction: weakly anharmonic qubit}
\label{sec:semi}

\begin{figure}[t]
\begin{center}
\includegraphics[width=0.39\textwidth]{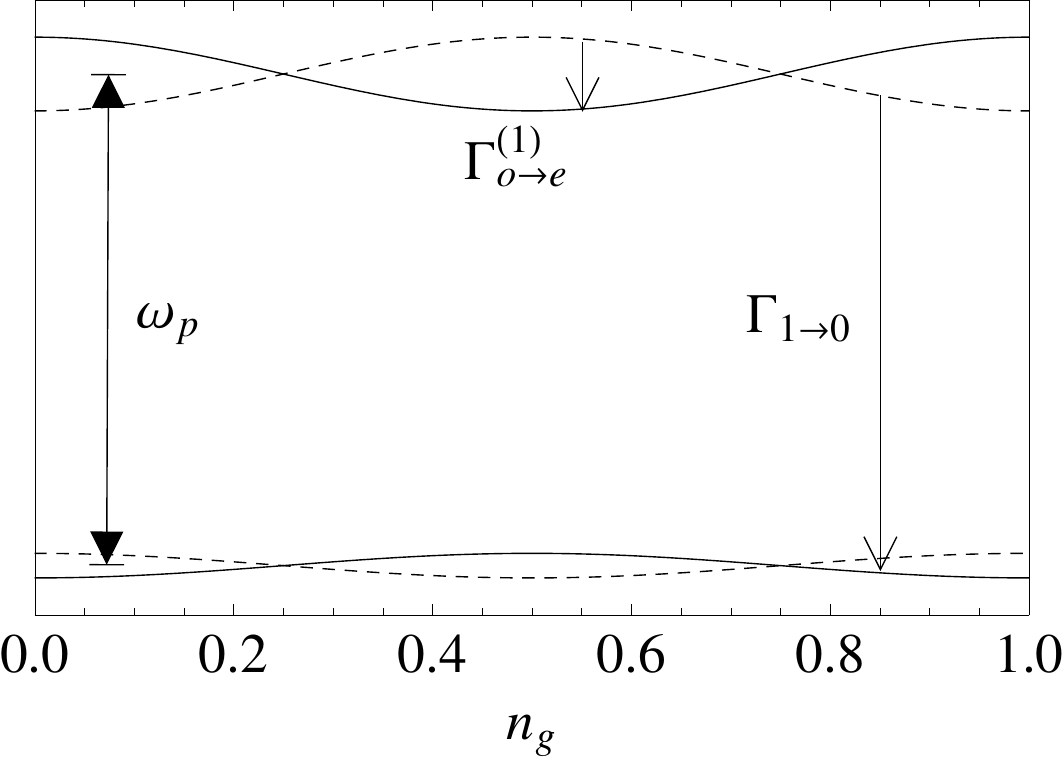}
\end{center}
\caption{Schematic representation of the transmon low energy spectrum as function of the dimensionless
gate voltage $n_g$. Solid (dashed) lines denotes even (odd) states (see also Sec.~\ref{sec:cpb}).
The amplitudes of the oscillations of the energy levels are exponentially small,\cite{transmon}
see Appendix~\ref{app:eosplit}; here
they are enhanced for clarity. Quasiparticle tunneling changes the parity of the qubit sate. The results
of Sec.~\ref{sec:semi} are valid for transitions between states separated by energy of the order of
the plasma frequency $\omega_p$, \eref{pl_fr}, and give, for example, the rate $\Gamma_{1\to 0}$.
For the transition rates between nearly degenerate states of opposite parity, such as $\Gamma^{(1)}_{o\to e}$,
see Appendix~\ref{app:eorate}.}
\label{fig:trans}
\end{figure}

As an application of the general approach described in the previous section, we consider
here a weakly anharmonic qubit, such as the transmon and phase qubits.
We start with the the semiclassical limit, i.e., we assume that the
potential energy terms in \eref{Hphi} dominate the kinetic energy term proportional to $E_C$.
This limit
already reveals a non-trivial dependence of relaxation on flux.
Note that assuming $E_L\neq 0$ we can eliminate $n_g$ in \eref{Hphi}
by a gauge transformation.\cite{flux_th} In the transmon we have
$E_L = 0$ and the spectrum depends on $n_g$, displaying both well
separated and nearly degenerate states, see Fig.~\ref{fig:trans}.
The results of this section can be applied to the single-junction
transmon when considering well separated states. The transition rate
between these states and the corresponding frequency shift are
dependent on $n_g$. However, since $E_C \ll E_J$ this dependence
introduces only small corrections to $\Gamma_{n\to n-1}$ and
$\delta\omega$; the corrections are exponential in $-\sqrt{8E_J/E_C}$. By
contrast, the leading term in the rate of transitions
$\Gamma_{e\leftrightarrow o}$ between the even and odd states is
exponentially small. The rate $\Gamma_{e\leftrightarrow o}$ of
parity switching is discussed in detail in Appendix~\ref{app:eorate}.

The potential energy in \eref{Hphi} is extremized at phase $\varphi_0$ satisfying
\be\label{extr}
E_J \sin \varphi_0 + E_L \left(\varphi_0 - 2\pi \Phi_e/\Phi_0\right) = 0.
\ee
For $E_J < E_L$ there is only one solution at the global minimum. For $E_J > E_L$ however,
there can be multiple minima; their number depends both on the ratio $E_J/E_L$ and the
external flux $\Phi_e$. Here we assume that the flux is such that distinct minima are not degenerate;
in particular, this means that the flux is tuned away from odd integer multiples of half the flux
quantum.\cite{noteEL} For the transmon with $E_L=0$, we can take $\varphi_0=0$ as solution to
\eref{extr}.
Next, we expand the potential energy around a minimum and find at quadratic order
\be\label{Hphi_2}
\hat{H}^{(2)}_\varphi = 4E_C \hat{n}^2 + \frac12 \left(E_L + E_J \cos \varphi_0\right)
\left(\hat{\varphi} - \varphi_0\right)^2 .
\ee
Fluctuations of the phase around $\varphi_0$ are small under the assumption
\be\label{small_fl}
n\frac{E_C}{\EZO} \ll 1,
\ee
where $n$ denotes the energy level and
\be\label{ezo}
\EZO = \sqrt{8 E_C \left(E_L + E_J \cos\varphi_0 \right)}
\ee
is the qubit frequency in the harmonic approximation.
Note that anharmonicity and quality factor $Q$
determine the operability of the system as a qubit.~\cite{Dev_rev} The anharmonic correction to the
transition frequencies can be calculated by considering the effect on the spectrum
of the next order in the expansion around $\varphi_0$ (cubic for the phase qubit, quartic for the transmon),
which defines an anharmonic potential well of finite depth $U$. Then the operability condition can be expressed
as $Q/n_w \gg 1$, where $n_w$ is the number of states in the potential well,
$n_w \sim U/\EZO$.\cite{note1} In a weakly anharmonic system,
$n_w$ can be large; however, if the quality factor is larger the system can be used as a qubit despite the
weak anharmonicity, as it is indeed the case for the transmon.~\cite{transmon}

The condition for small phase fluctuations in \eref{small_fl} enables us to calculate the matrix element
of operator $\sin \hat\varphi/2$ by expanding around $\varphi_0$ up to the second order
and using standard expressions for the matrix elements of the position operator between eigenstates
$|n\rangle$, $|m\rangle$
of the harmonic oscillator [cf. \eref{Hphi_2}]. To first order in $E_C/\EZO$ we find
(see also Appendix~\ref{app:home})
\be\label{harm_mat_el}\begin{split}
& \left|\langle m | \sin \frac{\hat\varphi}{2} | n\rangle \right|^2 =
\delta_{m,n}\left[1 - 2\frac{E_C}{\EZO}  \left(n + \frac12 \right)\right]\frac{1-\cos\varphi_0}{2} \\
& \hspace{1.4cm}
 + \frac{E_C}{\EZO} \left[ n\delta_{m,n-1} +  \left(n+1\right)\delta_{m,n+1}\right]\frac{1+\cos\varphi_0}{2} \, .
\end{split}\ee
Note that in the first term on the right hand side the corrections due to the non-linearity
of sine (the second term inside the square brackets) are indeed small if condition \rref{small_fl} is satisfied.
In addition, we have neglected here
the anharmonic corrections to the states used to calculate the matrix element; this is a good approximation
for low-lying levels $n \ll n_w$.\cite{note2} For the transmon ($\varphi_0=0$)
the leading term in \eref{harm_mat_el} is of linear order in $E_C/\EZO \ll 1$;
as we show in Appendix~\ref{app:tme} by including the first anharmonic correction to the states,
the next non-vanishing term in the square of the transmon matrix element is cubic in $E_C/\EZO$, rather
than quadratic as for the
harmonic oscillator. Therefore in the case of the transmon keeping only the leading term is a better
approximation than naively expected.

Equation \rref{harm_mat_el} shows that at leading order we can restrict our attention to transitions
involving only neighboring levels.
Concentrating here on low-lying levels,
using \esref{wif_gen},
\rref{S_ReY}, and \rref{harm_mat_el} we find the following relation between transition
rate and impedance
\be\label{GG_ReY}
\Gamma_{n \to n-1} - \Gamma_{n-1 \to n}= \frac{n}{C} \Re Y_\qp(\EZO) \frac{1+\cos\varphi_0}{2},
\ee
where we also used $E_C = e^2/2C$.
In the high-frequency regime, the upward transition rate can be neglected,
$\Gamma_{n-1 \to n} \simeq 0$, and the above expression simplifies to\cite{k_corr}
[see also \eref{S_ReY_hf} and the text that follows it]
\be\label{Gnn}\begin{split}
\Gamma_{n \to n-1} & = \frac{n}{C} \Re Y_\qp^{hf}(\EZO) \frac{1+\cos\varphi_0}{2} \\
& = n \frac{\omega_p^2}{\EZO} \frac{\ka}{2\pi} \sqrt{\frac{2\Delta}{\EZO}}
\left(1+\cos\varphi_0\right)\, .
\end{split}\ee
In the last expression we used \eref{ReYqp_hf} and introduced the plasma frequency
\be\label{pl_fr}
\omega_p = \sqrt{8E_C E_J}\, .
\ee
The above equation can also be obtained by substituting directly \eref{S_ReY_hf} into \eref{wif_gen}. For
$n=1$ and $\varphi_0 =0$ \eref{Gnn} reduces to the transition rate presented in \ocite{Martinis}.

The transition rate in \eref{Gnn} is proportional to the (possibly non-equilibrium) quasiparticle
density $\ka$ and depends on the external flux $\Phi_e$ via $\varphi_0$ and $\EZO$,
see \esref{extr} and \rref{ezo}.
The flux dependence is in general sensitive to the states involved in the transition.
This sensitivity can already be seen for transitions between harmonic oscillator states: due to
the non-linear interaction between phase and
quasiparticles, see \eref{HTle},
transitions between distant levels are possible. These transitions are suppressed by the smallness
of phase fluctuations when $E_C/\EZO \ll 1$. For example, the rate for the $2\to 0$ transition is
\be\label{G2}
\!\!\Gamma_{2\to 0} =
\frac{2\EZO}{\pi} \frac{1}{g_K} \Re Y_\qp^{hf}(2\EZO)\!
\left(\frac{E_C}{\omega_{10}}\right)^2\!
 \frac{1-\cos\varphi_0}{4}.
\ee
Note that in contrast to \eref{Gnn}, \eref{G2} cannot be written in terms of the real part of the total
admittance of the junction: while in \eref{Gnn} the phase enters via the factor
$(1+\cos\varphi_0)$ as in \eref{ReYJ},
in \eref{G2} $\Re Y_\qp$ is multiplied by $(1-\cos\varphi_0)$.
To obtain $\Gamma_{2\to 0}$ we substituted into \eref{wif_gen}
the high-frequency relation \rref{S_ReY_hf},
while the explicit form of the squared matrix element
$|\langle 0 |\sin (\hat\varphi/2) |2\rangle|^2$
is found by setting $n=2$, and keeping the leading term in $E_C/\EZO$, in the formula
\be\label{me_zn}
\left| \langle 0 |\sin \frac{\hat\varphi}{2}| n \rangle \right|^2 = e^{-\frac{E_C}{\omega_{10}}}
\left(\frac{E_C}{\omega_{10}}\right)^{n} \frac{1-(-1)^n\cos\varphi_0}{2n!}
\ee
derived in Appendix~\ref{app:home}. Equation \rref{me_zn} is valid for any ratio $E_C/\EZO$
for transitions between eigenstates of the harmonic oscillator.
When $\varphi_0=0$, \eref{me_zn} gives vanishing matrix elements for even $n$ -- this is an example of the
more general selection rule according to which only transitions between states of different parity are
allowed at $\varphi_0=0$.

The rate for transitions between excited states and
the ground state in the case of large phase fluctuations
can be obtained using \eref{me_zn} when $E_L \ll E_C$ and
$E_J \lesssim \omega_{LC} = \sqrt{8E_C E_L}$. The latter condition enables us to neglect
the Josephson energy term in \eref{Hphi}.
Then using \esref{wif_gen} and \rref{Sqp_hf} with $\omega_{if} = n \omega_{LC}$ we find
that the transition rate has a maximum for $n=n_0$ with $n_0 \approx E_C/\omega_{LC}$,
\be\label{Gnz}\begin{split}
\Gamma_{n \to 0} \simeq & \, \frac{\omega_p^2}{E_C} \frac{\ka}{2\pi} \sqrt{\frac{2\Delta}{E_C}}
\left[1-(-1)^n\cos 2\pi\Phi_e/\Phi_0\right] \\
& \times \frac{1}{\sqrt{2\pi n_0}}
\exp\left[-\frac{(n-n_0)^2}{2n_0}\right].
\end{split}\ee
Here we have
approximated $e^{-y} y^n/n!\sqrt{n} \simeq \exp[-(n-y)^2/2y]/\sqrt{2\pi}\, y$;
the approximation is valid for $y \gg 1$ and $|n-y| \lesssim \sqrt{2y}$.
Equation \rref{Gnz} shows that
when the charging energy is the dominant energy scale,
dissipation is the strongest for transitions between states
whose energy difference ($n \omega_{LC}$) corresponds to the energy
change ($E_C$) caused by the transfer of a single electron
through the barrier, as in the ``quasiparticle poisoning''
picture for the Cooper pair box.~\cite{lutchyn1} We
stress that in the present case charge is not quantized, due to the finite value of the inductive
energy $E_L$.\cite{flux_th}
We will comment on the relation between \eref{Gnz} and the transition rate in the Cooper pair box
in Sec.~\ref{sec:cpb}.

\subsection{Quality factor}
\label{sec:qf_s}

Returning now to the semiclassical regime of small $E_C$, \eref{Gnn} with $n=1$ enables us
to evaluate, in the high-frequency regime, the inverse $Q$-factor for the transition between the qubit states
\be\label{Q_single}
\frac{1}{Q_{10}}= \frac{\Gamma_{1\to 0}}{\EZO} =
\frac{1}{\pi g_K} \Re Y_\qp^{hf} (\EZO) \frac{E_C}{\EZO} \frac{1+\cos\varphi_0}{2} \, .
\ee
We stress that this formula is valid not only in thermal equilibrium, but also in the
presence of non-equilibrium quasiparticles with characteristic energy $\delta E \ll \EZO$.
We can generalize \eref{Q_single} to account for the possible coexistence of non-equilibrium and
thermal quasiparticles. We take the distribution function in the form
\be\label{f_eq_ne}
f_E(\epsilon) = f_{ne} (\epsilon) + f_{eq} (\epsilon) \, ,
\ee
where $f_{ne}$ is the non-equilibrium contribution, insensitive to temperature and satisfying the
high-frequency condition
$\EZO \gg \delta E$, and $f_{eq}$ is the equilibrium distribution of \eref{f_th}. Noting that
within our assumption the two terms in $f_E$ contribute separately to the transition rates and
that for the thermal part we cannot in general neglect the ``upward'' transitions,
using \esref{wif_gen}, \rref{Sqp_th}, \rref{Sqp_hf}, and \rref{harm_mat_el} we find
\be\label{Q1_nt}\begin{split}
\frac{1}{Q_{10}}= \frac{\Gamma_{1\to 0} + \Gamma_{0\to 1}}{\EZO} =
\frac{1+\cos\varphi_0}{2\pi} \frac{\omega_p^2}{\EZO^2}
\bigg[
x_{ne}\sqrt{\frac{2\Delta}{\EZO}}  & \\ + 4
e^{-\Delta/T} \cosh\left(\frac{\EZO}{2 T}\right) K_0 \left(\frac{\EZO}{2 T}\right)
\bigg], &
\end{split}\ee
where $x_{ne}$ is the normalized non-equilibrium quasiparticle density
[cf. \eref{xqp_def}].

Recently good agreement between theory, \eref{Q1_nt}, and experiment has been shown
for single-junction transmons ($\varphi_0=0$, $\EZO = \omega_p$) in the temperature range
10-210~mK.~\cite{paik} However, while these measurements indicate that thermal quasiparticles
are the main cause of relaxation above $\sim 150$~mK, one cannot
conclude that non-equilibrium quasiparticles are present from the lower temperature data:
by Matthiessen rule, any other relaxation mechanism which is independent of (or weakly dependent on)
temperature would have the same limiting effect on $Q_{10}$ as the first term in square brackets
in \eref{Q1_nt}. As we will discuss in more detail in Sec.~\ref{sec:split_tr},
similar measurements on a flux-sensitive device should enable one to decide on the presence
of non-equilibrium quasiparticles, since \eref{Q1_nt} [and its analogous for the split transmon,
\eref{Qqp_m}]
describes the effect of flux on both equilibrium
and non-equilibrium quasiparticle contributions to $Q_{10}$,
and other sources of relaxation respond differently to the flux.

\subsection{Frequency shift}
\label{sec:freq_sh}

A further test of the theory presented in Sec.~\ref{sec:th_s} is provided by the measurement
of the qubit resonant frequency. In the semiclassical regime of small $E_C$,
the qubit can be described by the effective circuit
of Fig.~\ref{fig1}(b), with the junction admittance $Y_J$ of \eref{YJ}, $Y_C= i\omega C$, and
$Y_L = 1/i\omega L$ [the inductance is related to the inductive energy by $E_L = (\Phi_0/2\pi)^2/L $].
As discussed in \ocite{prl}, for parallel
elements the total admittance $Y$ is the sum of their admittances,
\be
Y = Y_J + Y_C +Y_L \, ,
\ee
and the resonant frequency $\omega_r$ is the zero of the total admittance,
$Y(\omega_r)=0$. In the absence of quasiparticles we find $\omega_r = \EZO$ with $\EZO$
of \eref{ezo}.

In the presence of quasiparticles, by considering their effect on the junction admittance at linear
order in the quasiparticle
density $\ka$ and Andreev level occupation $\kaA$ we obtain
\be
\omega_r = \EZO + \delta \omega
\ee
with
\be\label{do_circ}\begin{split}
\delta \omega = \frac{i}{2C}Y_\qp(\EZO)\frac{1+\cos \varphi_0}{2}
-\frac{\pi g_T \Delta}{C \EZO}\kaA \cos\varphi_0 \\ -
\frac{\pi g_T \Delta}{2C \EZO}\ka \cos\varphi_0 \, .
\end{split}\ee
The last term in \eref{do_circ} originates from the gap suppression by quasiparticles [cf. \eref{delqp}].
This term was neglected in \ocite{prl} as it is subleading in the high-frequency regime considered there
[see \eref{do_hf}]. The correction $\delta\omega$ has both real and imaginary parts. The imaginary part
coincides\cite{prl} with half the dissipation rate in \eref{Gnn} for the $n=1\to 0$ transition.
Here we show that the real part of $\delta\omega_r$ obtained
in the effective circuit approach agrees with the quantum mechanical calculation.

Within the harmonic approximation of \eref{Hphi_2},
the energy difference $\omega_i$ between the neighboring levels $E_{i+1}$ and $E_i$,
\be
\omega_i \equiv E_{i+1} - E_i = \EZO \, ,
\ee
is of course independent of the level index $i$. The quasiparticle corrections to energy levels
of Sec.~\ref{sec:encorr} cause a correction $\delta\omega_i$ to $\omega_i$,
\be\label{dom}
\delta \omega_i = \delta E_{i+1} - \delta E_i \, .
\ee
As we show below, at leading order in $E_C/\EZO$ this correction is also independent of level index, i.e,
it represents a renormalization of the system resonant frequency.

As in \eref{de_sep}, we separate the contributions due to change in the Josephson energy and due to quasiparticle
tunneling,
\be
\delta\omega_i = \delta\omega_{i,E_J} + \delta\omega_{i,\qp}.
\ee
For the first term on the right hand side, we use \eref{de_EJ}
together with the matrix element of $\cos\hat\varphi$
at first order in $E_C/\EZO$  [see \eref{me_cos}],
\be\label{cos_me}
\langle i|\cos\hat\varphi|i \rangle \simeq \cos\varphi_0 \left[1-\frac{4E_C}{\EZO}\left(i+\frac12\right)\right],
\ee
to find
\be
\delta\omega_{i,E_J} = -\frac12 \frac{\omega_p^2}{\EZO} \cos\varphi_0 \left(\ka +2\kaA\right) \, .
\ee
As discussed in Sec.~\ref{sec:encorr}, the term proportional to $\ka$ is due to the gap
suppression in the presence of quasiparticles, \eref{delqp}, while $\kaA$ accounts for the
occupation of the Andreev bound states.

For the quasiparticle tunneling term, we substitute \eref{harm_mat_el} into \eref{de_qp} to get
\be
\delta\omega_{i,\qp} = \frac{E_C}{\EZO} \left[F_\qp(\EZO) + F_\qp(-\EZO) \right]
\frac{1+\cos\varphi_0}{2} \, .
\ee
Finally, using the relation \rref{F_ImY} and adding the two terms we arrive at
\be\label{do_f}\begin{split}
\delta\omega_i = & -\frac{1}{2C} \Im Y_\qp(\EZO) \frac{1+\cos\varphi_0}{2} \\
& -\frac12 \frac{\omega_p^2}{\EZO} \cos\varphi_0 \left(\ka +2\kaA\right)\, .
\end{split}\ee
This expression agrees with the real part of \eref{do_circ}. We note that by extending the above
consideration to include the next order
in $E_C/\EZO$, anharmonic corrections to the spectrum can be calculated. They are dominated by the
anharmonicity of the cosine potential in \eref{Hphi}, with quasiparticles contributing negligible
additional corrections. For the case of the transmon, the leading anharmonicity can be found in
\ocite{transmon}.

In the high-frequency regime, using \eref{Y2_hf} the relative frequency shift is
\be\label{do_hf}\begin{split}
\frac{\delta\omega_i}{\EZO} = & \frac12 \frac{\omega_p^2}{\EZO^2} \Bigg[\kaA \left(1-\cos\varphi_0\right)
\\ & -\ka
\left(\frac{1+\cos\varphi_0}{2\pi}\sqrt{\frac{2\Delta}{\EZO}}+\cos\varphi_0\right)\Bigg].
\end{split}\ee
Note that in the limit $\EZO \ll \Delta$ we can neglect the cosine compared to the term
multiplied by square root inside
round brackets. However, this cosine term is the appropriate subleading contribution, since the terms neglected
in deriving the energy corrections presented in Sec.~\ref{sec:encorr} are suppressed
by $\EZO/\Delta$ with respect to the leading contribution.

In recent experiments with single-junction
transmons~\cite{paik} relative shifts of order $10^{-5}$ have been measured at temperatures $\sim 200$~mK,
in agreement with \eref{do_f}. Together with the above mentioned measurements of the transition rates in the
same devices, this is an additional, independent check of the validity
of the present theory in the regime $T\gtrsim 150$~mK.
While in the transmon ($\varphi_0 =0$) there are no Andreev bound states
[indeed, in this case their contribution to the frequency shift is absent, see \eref {do_hf}],
in a phase qubit both Andreev levels occupation $\kaA$ and free quasiparticle density $\ka$ affect the frequency.
Assuming that the two quantity are proportional, $\kaA \propto \ka$ ,the ratio between frequency
shift, \eref{do_hf}, and transition rate, \eref{Gnn}, in the high frequency regime is independent of the
quasiparticle density. The constancy of this ratio has been recently verified by injecting a variable
(but unknown) number quasiparticles in a phase qubit.~\cite{martinis2}

\section{Single junction: strong anharmonicity}
\label{sec:further}

Here we consider the regime, complementary to that of the previous section,
of qubits with large anharmonicities. We study first the single junction Cooper pair box (CPB);
as for the transmon, it is insensitive to flux, but in contrast to the transmon the CPB properties are
strongly affected by the value of the dimensionless gate voltage $n_g$.
Then we analyze a flux qubit, for which the external flux is tuned near half the flux
quantum, $\Phi_e \approx \Phi_0/2$.

\subsection{Cooper pair box}
\label{sec:cpb}

The CPB is described by \eref{Hphi} with $E_L=0$ and $E_C \gg E_J$. In this limit, it
is convenient to rewrite the Hamiltonian in the charge basis as~\cite{lutchyn1}
\be\begin{split}
\hat H = &  \, E_C \sum_q  \Big(q - 2n_g\Big)^2 |q\rangle\langle q | \\  & -
\frac{1}{2}E_J \sum_q \Big(|q\rangle\langle q+2| + |q+2\rangle\langle q| \Big).
\end{split}\ee
The eigenstates have definite parity (even/odd) and are given by linear combinations of
even/odd charge states. The CPB operating point is, without loss of generality, at $n_g=1/2$.
Near this operating point, the CPB is well described by the reduced Hamiltonian
\be\label{HCPB}
H_{CPB} = \left(\begin{matrix}
E_C (2n_g)^2 & 0 & -E_J/2 \\
0 & E_C (2n_g-1)^2 & 0 \\
-E_J/2 & 0 & E_C (2n_g -2)^2
\end{matrix}\right).
\ee

\begin{figure}
\includegraphics[width=0.42\textwidth]{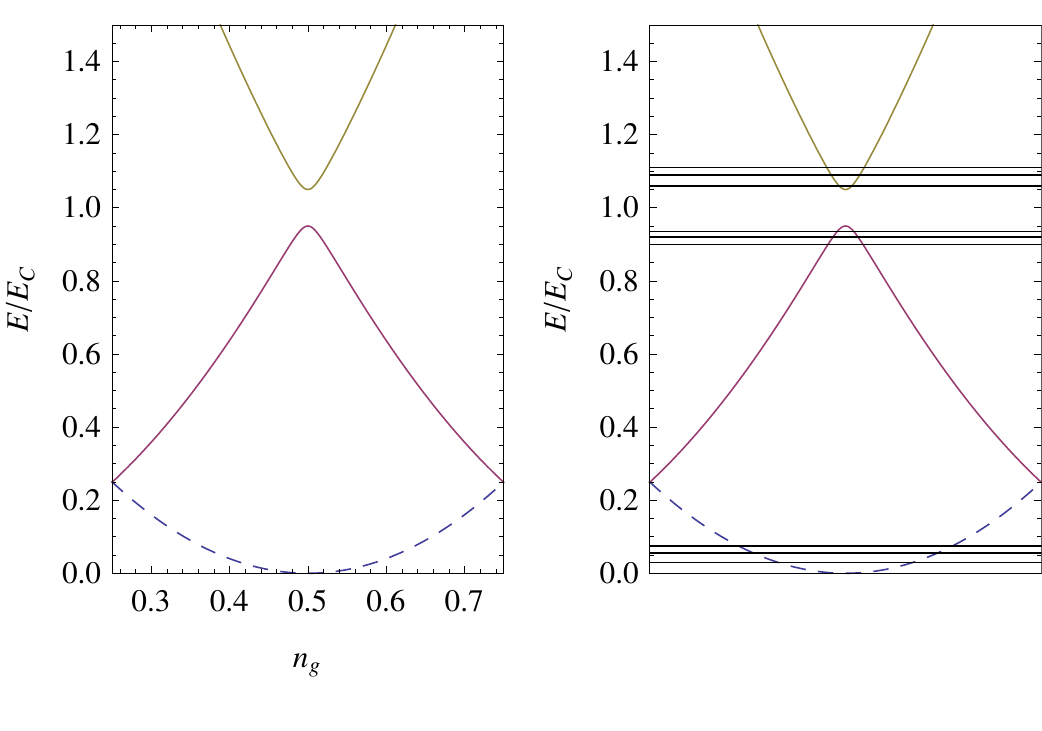}
\caption{Left panel: spectrum of the reduced CPB Hamiltonian, \eref{HCPB}, around the operating point $n_g=1/2$ for
$E_J = 0.1 E_C$.
Dashed line: energy of the odd state, \eref{en_odd}. Solid lines: energies of ground (bottom) and excited
(top) even states, \eref{en_even}. Right panel: in the presence of a small inductive energy $E_L$, the CPB
bands act as
potentials in the quasimomentum space, see \ocite{flux_th}. Dense horizontal lines represent a few energy levels
near the edges of the bands.}
\label{fig:CPB}
\end{figure}

The reduced CPB Hamiltonian has a single odd eigenstate, the $|q=1\rangle$ charge
state,
\be\label{st_o}
|o,0;n_g\rangle = |1\rangle ,
\ee
with $n_g$-dependent eigenenergy
\be\label{en_odd}
E_0 (n_g) = E_C (2n_g-1)^2 \, ,
\ee
and two even eigenstates, $|e,\pm;n_g\rangle$, with energies
\be\label{en_even}
E_\pm (n_g) = E_C+E_0(n_g) \pm \frac12\EZO(n_g)\, .
\ee
The qubit frequency depends on the gate voltage as
\be
\EZO(n_g) = \sqrt{(4E_C)^2 (2n_g-1)^2+ E_J^2} \, .
\ee
Note that at the operating point we have $\EZO(1/2) = E_J$ and that the frequency rises quickly
at a narrow distance from the optimal point, more than doubling for
$|n_g-1/2| \sim E_J/E_C \ll 1$.
In terms of the charge states, the two even eigenstates are
\be\label{st_e}\begin{split}
|e,-;n_g\rangle & = \cos \theta |0\rangle + \sin \theta |2\rangle \, , \\
|e,+;n_g\rangle & = \sin \theta |0\rangle - \cos \theta |2\rangle \, ,
\end{split}\ee
where
\be\label{cos_th}
\cos \theta = \frac{1}{\sqrt{2}}\sqrt{1-\frac{4E_C (2n_g-1)}{\EZO(n_g)}}.
\ee

The non-vanishing matrix elements of $\sin \hat\varphi/2$ can be readily obtained using the charge basis
form of this operator
\be\label{sin_cb}
\sin \frac{\hat\varphi}{2} = \frac{1}{2i}\sum_q \Big(|q+1\rangle\langle q| -|q\rangle\langle q+1| \Big).
\ee
For the states in \esref{st_o} and \rref{st_e} we find
\be\label{cpb_me}
\left|\langle o,0;n_g |\sin\frac{\hat\varphi}{2} |e,\pm; n_g \rangle\right|^2 =
\frac{1}{4}\left[1 \pm \frac{E_J}{\EZO(n_g)}\right].
\ee
We stress that the transitions are not between the qubit (i.e., even) states, but between
the even and odd states; the corresponding transition frequencies are
$\omega_\pm(n_g) = E_C \pm \EZO(n_g)/2$, see \esref{en_odd} and \rref{en_even}.
Therefore the tunneling of a quasiparticle into the CPB changes the parity of the state, an
effect known as ``quasiparticle poisoning''.~\cite{matveev}
Substituting the matrix element \rref{cpb_me} into \eref{wif_gen}
and using the high-frequency expression \rref{Sqp_hf} we find
\be\label{Gcpb}
\Gamma_{e,+ \to o,0} = \left[1 + \frac{E_J}{\EZO(n_g)}\right] \frac{2E_J}{\pi} \, \ka
\sqrt{\frac{2\Delta}{\omega_+(n_g)}}
\ee
for the transition between even excited and odd states.
In thermal equilibrium with $T\ll \omega_+(n_g)$, using \eref{nqp_th} we obtain
\be\label{G_lut}
\Gamma_{e,+ \to o,0} = \left[1 + \frac{E_J}{\EZO(n_g)}\right] \frac{4E_J}{\sqrt{\pi}}
\sqrt{\frac{T}{\omega_+(n_g)}} e^{-\Delta/T}.
\ee
Within our approximations, this expression reproduces
(after implementing the corrections described in \ocite{erratum} and up to a numerical prefactor)
the decay rate calculated in \ocite{lutchyn1}
for the ``open'' qubit at the operating point $n_g=1/2$. For the
transition between even ground and odd states the matrix element in \eref{cpb_me}
vanishes at the operating point. This vanishing is
a consequence of the low-energy approximation that lead to \eref{HTle}: as the results
of Refs.~\onlinecite{lutchyn1,lutchyn2} show, the contributions that we neglect cause a
finite transition rate, which is
suppressed by a small factor of order $E_C/2\Delta$ in comparison with the transition rate from
even excited to odd state.

We note that while in all the above expressions the distance $|2n_g-1|$ from the operating point can
be large compared to the small parameter $E_J/E_C \ll 1$, the description based on \eref{HCPB} is valid if
other charge states can
be neglected, which limits the range of validity to $|2n_g-1|<1/2$ (with $|2n_g-1| - 1/2 \gg E_J/E_C$).
For example, at $2n_g-1 \simeq 1/2$ the charge states
$|0\rangle$ and $|3\rangle$ are nearly degenerate and we can expect an enhanced transition rate
$\Gamma_{e,+ \to o,3}$ in comparison to the rate $\Gamma_{e,+ \to o,0}$ that we have considered above.

Finally, let us comment on the relationship between the transition rate in the CPB and in the
inductively shunted Josephson junction with large charging energy [see the paragraph containing
\eref{Gnz}]. As shown schematically in the right panel of Fig.~\ref{fig:CPB} and discussed in detail
in \ocite{flux_th}, the spectra of the two systems are distinct even
in the limit of small inductive energy $E_L$: in the CPB ($E_L = 0$) the energy levels form bands as $n_g$
varies, while for any non-zero $E_L$ the gate voltage $n_g$ can be ``gauged away'' and the
spectrum consists of discrete levels that become denser as $E_L$ decreases.
Despite these differences, the ac responses of the two systems due to charge coupling
agree in this limit.\cite{flux_th} Similarly, we now show agreement for the quasiparticle transition rates.
We note that when
taking the limit $E_L \to 0$, the condition $E_J\lesssim \omega_{LC}$ for the validity of \eref{Gnz}
for the rate $\Gamma_{n\to 0}$
requires that we also take $E_J \to 0$.~\cite{cpbcomp} Moreover, since the final state considered
in deriving the rate $\Gamma_{n\to 0}$ is the lowest possible state, the corresponding final state
in the CPB is
either the even ground state at $n_g=0$ or the odd ground state at $n_g =1/2$. Indeed, the
width of the ground state (in quasimomentum space -- see Fig.~\ref{fig:CPB} and \ocite{flux_th}) is
$\propto (E_L/E_C)^{1/4}$, so that as $E_L\to 0$ the state is localized at the bottom of the band.
Note that following
the same procedure detailed above it is straightforward to show that the transition rate
$\Gamma_{o,+ \to e,0}$ at $n_g = 0$ coincides with $\Gamma_{e,+ \to o,0}$ at $n_g=1/2$; hence for
our purposes the two possibilities are equivalent.
At finite $E_L$, the total transition rate to the ground state is obtained by summing \eref{Gnz} over all initial
levels $n$. Due to the Gaussian factor in the second line of \eref{Gnz}, the number of levels that contribute
to the total rate is approximately $\sqrt{n_0} \propto (E_C/E_L)^{1/4}$, which grows as the inductive
energy diminishes. However, the energy of the contributing levels tends to the charging energy,
as can be seen by
rewriting identically the argument in the exponential of the Gaussian factor
as $-(E_n-E_C)^2/2E_C \omega_{LC}$, where $E_n = n\omega_{LC}$; this agrees with
frequency for the $e,+ \to o,0$ transition at $n_g=1/2$ in the CPB being approximately
$E_C$ in the small $E_J$ limit.
Using \eref{Gnz}, performing the sum over levels, and taking the limit $E_L \to 0$, we find
\be\label{Gec}
\lim_{E_L \to 0} \sum_n \Gamma_{n \to 0} = \frac{4E_J}{\pi}\, \ka
\sqrt{\frac{2\Delta}{E_C}}\, ,
\ee
which coincides with the leading term of \eref{Gcpb} in the limit $E_J \to 0$ at the operating point $n_g=1/2$.

\subsection{Flux qubit}
\label{sec:fl_qb}

\begin{figure}
\begin{center}
\includegraphics[width=0.43\textwidth]{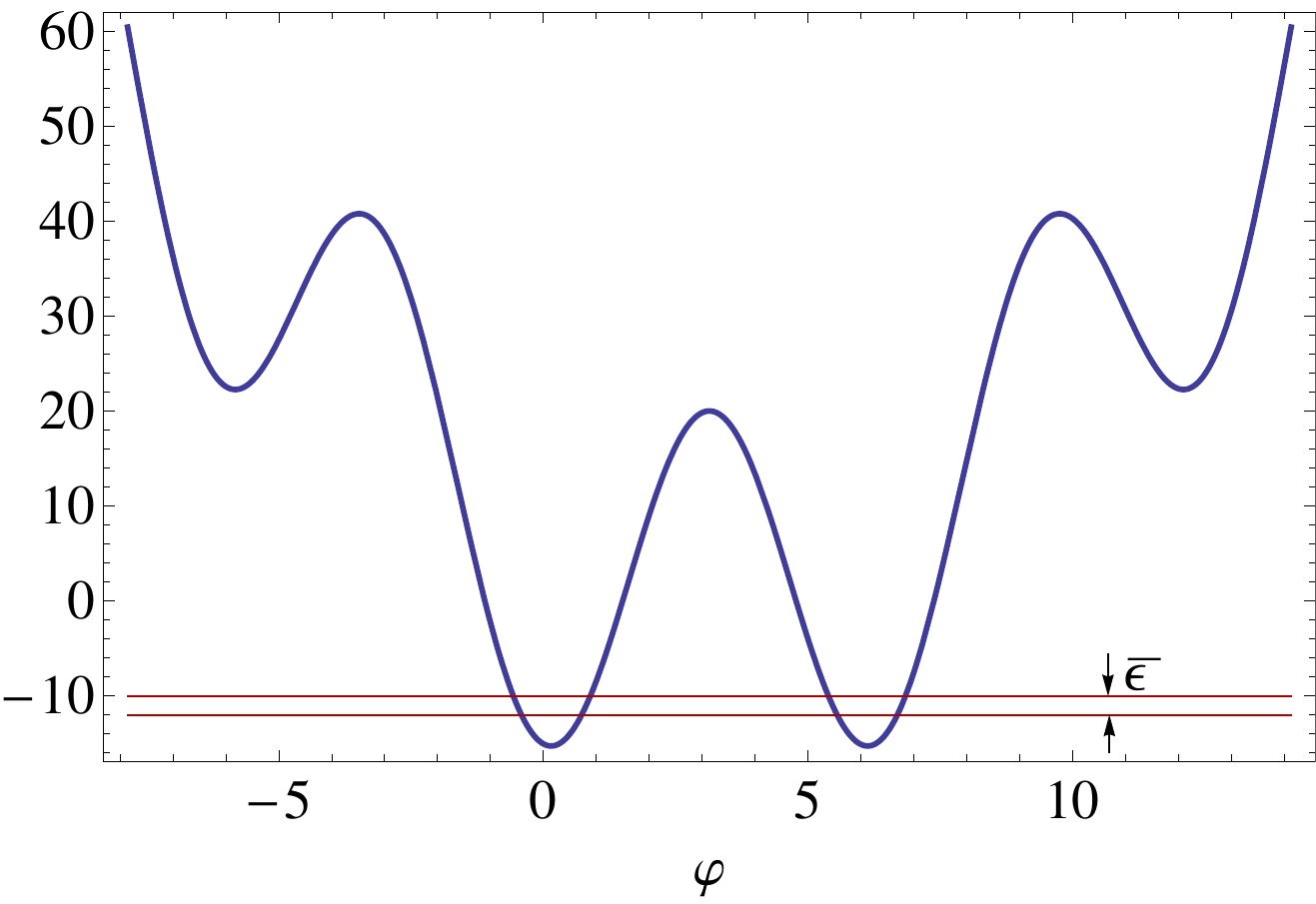}
\end{center}
\caption{Potential energy (in units of $E_L$) for a flux qubit biased at $\Phi_e = \Phi_0/2$ with
$E_J/E_L=10$. The horizontal
lines represent the two lowest energy levels, with energy difference $\bar\epsilon$ given in \eref{e0_eff}.}
\label{fig:fl_q}
\end{figure}

As a second example of a strongly anharmonic system, we consider here a flux qubit, i.e.,
in \eref{Hphi} we assume $E_J > E_L$ and take
the external flux to be close to half the flux quantum, $\Phi_e \approx \Phi_0/2$.
Then the potential has a
double-well shape and the flux qubit ground states $|-\rangle$ and excited state
$|+\rangle$ are the lowest tunnel-split eigenstates in this
potential,\cite{Dev_rev} see Fig.~\ref{fig:fl_q}.
The non-linear nature of the $\sin \hat\varphi/2$ qubit-quasiparticle coupling in \eref{HTle}
has a striking effect on the transition rate
$\Gamma_{+\to -}$, which vanishes at
$\Phi_e=\Phi_0/2$ due to destructive interference: for flux biased at half the flux quantum
the qubit states $|-\rangle$, $|+\rangle$ are respectively symmetric and antisymmetric
around $\varphi=\pi$, while
the potential in Eq.~(\ref{Hphi}) and the function $\sin \varphi/2$ in
Eq.~(\ref{wif_gen}) are symmetric. Note that the latter symmetry and its consequences are absent in
the environmental approach in which a linear
phase-quasiparticle coupling is assumed.

Analytic evaluation of the matrix element determining the transition rate [\eref{wif_gen}]
at finite $\Phi_e-\Phi_0/2$ is possible when $E_C \ll E_J$ and
the tunnel splitting $\bar\epsilon$ is small compared to inductive and plasma
energies, $\bar\epsilon \ll 2\pi^2 E_L \ll \omega_p $;
an estimate for the splitting is given
below in \eref{e0_eff}. With the above assumptions we can use
a tight-binding approach. Neglecting tunneling the wavefunctions $|m\rangle$ are, as a first approximation,
ground state wavefunctions of the harmonic oscillator with
frequency $\omega_p$ and oscillator length $\ell = \sqrt{8E_C/\omega_p}$ localized around the (flux-dependent)
minima $\varphi_m$ of the potential energy,
\be\label{loc_st}
\langle \varphi | m\rangle  = \left(\frac{1}{\pi \ell^2}\right)^{1/4}
e^{-(\varphi - \varphi_m)^2/2\ell^2}.
\ee
The minima are found by solving \eref{extr} approximately, using the condition $E_L \ll E_J$
(which follows from the above assumptions) to get
\be\label{min_pos}
\varphi_m \simeq 2\pi \left[m - \frac{E_L}{E_J}(m-f)\right] \, , \qquad f = \frac{\Phi_e}{\Phi_0}\, .
\ee
The energies of the localized states are (up to a constant term)
\be\label{en_loc}
E_m = 2 \pi^2 \bar{E}_L (m-f)^2 \, ,
\ee
where
\be
\bar{E}_L = E_L \left(1-\frac{1}{\beta}\right) \, , \qquad \beta = \frac{E_J}{E_L}
\ee
takes into account corrections small in $1/\beta \ll 1$.
The above results are valid for $|m E_L/E_J| \ll 1$.
Still neglecting tunneling, the matrix element of $\sin\hat\varphi/2$
between states localized in different wells
vanishes, but the diagonal matrix element is finite due to the shift of the minima away from $2\pi m$, see
\eref{min_pos}. Using the states in \eref{loc_st} we obtain
\be\label{mef0}
\langle j | \sin \frac{\hat\varphi}{2} |m \rangle \simeq - (-1)^m \pi \frac{E_L}{E_J} (m-f) \delta_{m,j} \, .
\ee

To include the effect of tunneling we allow for the possibility of transitions between neighboring
wells with amplitude $\bar\epsilon/2$.
As we are interested in the two lowest eigenstates for $f$ near 1/2,
we consider only the $m=0,1$ wells and
the effective Hamiltonian has the form
\be
\hat H =  \left(\begin{matrix}
2\pi^2 \bar{E}_L f^2 & -\bar\epsilon/2 \\
-\bar\epsilon/2 & 2\pi^2 \bar{E}_L(1-f)^2
\end{matrix}\right).
\ee
The eigenenergies are [cf. \esref{en_even}-\rref{cos_th}]
\be
E_\pm(f) = \frac{\pi^2}{2} \bar{E}_L \left[1+(2f-1)^2\right] \pm \frac12 \EZO(f)
\ee
with the flux-dependent qubit frequency
\be\label{o_f}
\EZO(f) =  \sqrt{\bar\epsilon^2 + \left[(2\pi)^2\bar{E}_L (f-1/2)\right]^2}\, ,
\ee
while the eigenstates are
\be\label{f_s}\begin{split}
|- \rangle & = \cos \theta |0\rangle + \sin \theta |1\rangle \, ,\\
|+ \rangle & = \sin \theta |0\rangle - \cos \theta |1\rangle \, ,
\end{split}\ee
with
\be\label{de_f}
\cos\theta = \frac{1}{\sqrt{2}}\sqrt{1-\frac{(2\pi)^2 \bar{E}_L (f-1/2)}{\EZO(f)}}\, .
\ee
The tunnel splitting $\bar\epsilon$ entering in the above formulas can be estimated
by noting that
due to the assumption $\beta \gg 1$
the wells are nearly symmetric. Neglecting the asymmetry [i.e., considering the potential in \eref{Hphi}
at $f=1/2$], the width and height of the tunnel barrier are approximately $2\pi(1-1/\beta)$
and $2\bar{E}_J$, respectively, with
\be
\bar{E}_J = E_J \left[1-\frac{\pi^2}{4}\frac{1}{\beta}\left(1-\frac{1}{\beta}\right)\right].
\ee
To account for the height and width at $E_L\neq 0$,
we treat the two wells as cosine potentials with renormalized coefficients. That is, we consider
each well to be described by the Hamiltonian given in \eref{H_w} with the substitutions $E_J \to \bar{E}_J$ and
$E_C \to \bar{E}_C$, where
\be
\bar{E}_C = E_C\frac{1}{\left(1-1/\beta\right)^2}.
\ee
Then we can use the known asymptotic formula~\cite{transmon,flux_th,jcp}
for the splitting $\epsilon_0$ in the periodic cosine potential
(i.e., for $E_L=0$; see Appendix~\ref{app:eosplit} for a derivation of this formula)
\be\label{ep0_def}
\epsilon_0 = 4\sqrt{\frac{2}{\pi}} \, \omega_p \left(\frac{8E_J}{E_C}\right)^{1/4}
e^{-\sqrt{8E_J/E_C}}
\ee
to find
\be\label{e0_eff}
\bar\epsilon = 2\sqrt{\frac{2}{\pi}} \sqrt{8 \bar{E}_{J} \bar{E}_{C}}
\left(\frac{8\bar{E}_{J}}{\bar{E}_{C}}\right)^{1/4}
e^{-\sqrt{8\bar{E}_{J}/\bar{E}_{C}}}.
\ee
Here the numerical prefactor is smaller by factor of 2 in comparison with \eref{ep0_def}
to account for tunneling being between two wells rather than
in a periodic potential.~\cite{jcp}

Turning now to the matrix element $\langle j | \sin\hat\varphi/2 | m \rangle$,
the diagonal elements $j=m=0, \, 1$ are still
approximately given by \eref{mef0}. Tunneling introduces finite but exponentially small
off-diagonal elements which, similarly to the splitting, can be calculated using the semiclassical approximation.
Using the wavefunctions derived in Appendix~\ref{app:eosplit} we arrive at [cf. \eref{sfin}]
\be\label{meov}
\langle 1 | \sin \frac{\hat\varphi}{2} |0 \rangle
\simeq D \left(\frac{\bar{E}_J}{\bar{E}_C}\right)^{1/3}
\frac{\bar\epsilon}{2\sqrt{2} \bar{E}_J}
\ee
with $D\approx 1.45$, see \eref{Ddef}.
We can now calculate the matrix element of $\sin\hat\varphi/2$ between qubit states $|\pm\rangle$
in \eref{f_s}
using \esref{mef0} and \rref{meov} to obtain
\be\label{fl_qb_me}\begin{split}
\langle - | \sin \frac{\hat\varphi}{2} |+\rangle =
\pi(f-1/2)\frac{\bar\epsilon}{\EZO(f)}
\qquad \\ \times
\left[\frac{E_L}{E_J}+\sqrt{2}\pi D \frac{\bar{E}_L}{\bar{E}_J}
\left(\frac{\bar{E}_J}{\bar{E}_C}\right)^{1/3}\right].
\end{split}\ee
Here the first term in square brackets is the combination of the two intrawell contributions [\eref{mef0}]
while the second one originates from the under-barrier tunneling [\eref{meov}].
Comparing \eref{fl_qb_me} to numerical calculations, we find that
near half the flux quantum, $|f-1/2|\lesssim \bar\epsilon/2\pi^2 E_L$,
the two approaches give the same dependence on flux and agree on the order
of magnitude of the matrix element, with \eref{fl_qb_me} providing a smaller estimate than the numerics
by a factor of about $2/3$.
For $|f-1/2|\gtrsim \bar\epsilon/(2\pi)^2 E_L$ the flux dependence in \eref{fl_qb_me}
via the factor $(f-1/2)/\EZO(f)$ can be neglected and the right hand side reduces to a flux-independent constant.
However, this behavior is an artifact of our approximations: for these larger deviations
of flux from half the flux quantum the matrix element acquires additional flux dependence, beyond
that given in \eref{fl_qb_me}, once the asymmetry of the potential is taken into account.
Moreover, for very small flux, $|f| \lesssim (\bar\epsilon/4\sqrt{2}\pi^2 E_L)^2$, mixing
of the sate localized in well $m=1$ with that
localized in well $m=-1$ cannot be neglected and the matrix element has a narrow peak around zero flux.
Substituting \eref{fl_qb_me} into \eref{wif_gen}, keeping the leading contribution, and using the
relation \rref{S_ReY_hf}, we find for the transition rate in the high-frequency regime\cite{mf}
\be\label{G_fl_q}\begin{split}
\Gamma_{+ \to -} = \frac{\EZO}{\pi} \frac{1}{g_K} \Re Y^{hf}_\qp (\EZO)
\left(\frac{\bar\epsilon}{4\pi \bar{E}_J}\right)^2
\left(1 - \frac{\bar\epsilon^2}{\EZO^2}\right)
\\ \left(\sqrt{2}\pi D\right)^2 \left(\frac{\bar{E}_J}{\bar{E}_C}\right)^{2/3}
\end{split}\ee
with $\Re Y^{hf}_\qp$ of \eref{ReYqp_hf}.

The rate in \eref{G_fl_q} depends on reduced flux $f$ via the qubit frequency, see \eref{o_f}. In particular,
for external flux equaling half the flux quantum we have $\EZO(1/2) = \bar\epsilon$ and
the transition rate vanishes, as discussed above.
In the previous section we mentioned in the text after \eref{G_lut} that for the Cooper pair
box the vanishing of the rate at the operating point is valid up to small corrections, being a
consequence of the low-energy approximation for the tunneling Hamiltonian in \eref{HTle}. The same
is true for the flux qubit; in the present case, the parameter suppressing these corrections is
exponentially small, being given by $\bar\epsilon/2\Delta$. Note that if keeping in \eref{HT} the
contributions beyond the low energy approximation, the operators accounting for the qubit-quasiparticle
interaction cannot be reduced to $\sin\hat\varphi/2$; therefore, for these additional contributions
the symmetry argument given at the beginning
of this section for the vanishing of the transition rate at $f=1/2$ does not hold.

\section{Multiple-junction qubits: general theory and applications}
\label{sec:multi}

\begin{figure}
\begin{center}
\includegraphics[width=0.39\textwidth]{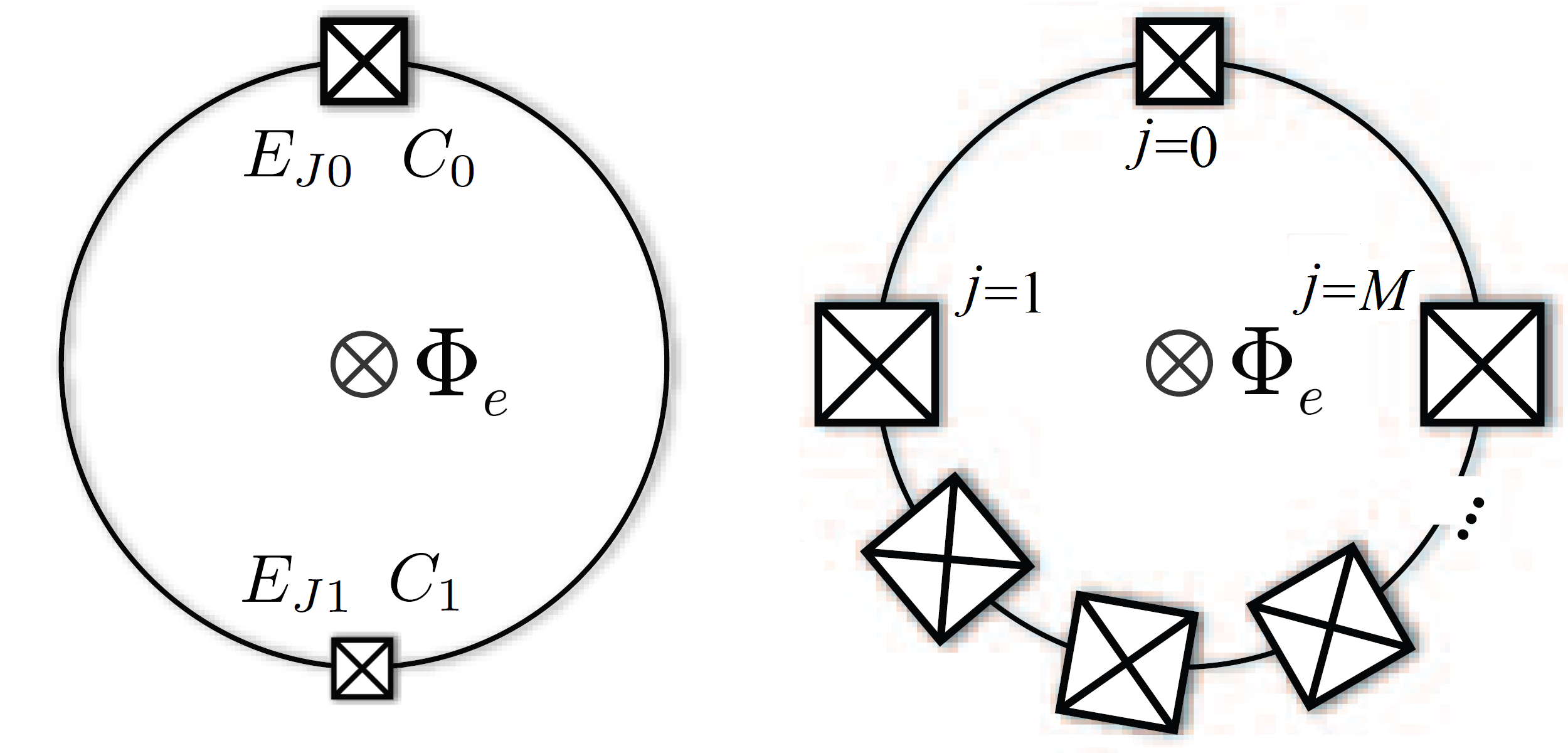}
\end{center}
\caption{Left: schematic representation of the split transmon with two (possibly different) junctions.
Right: in the fluxonium a weaker junction ($j=0$) is connected to a large junction array
($j=1,\ldots,M$).}
\label{fig:sp_tr}
\end{figure}

In this section we generalize the theory of Sec.~\ref{sec:th_s} to the case of
systems containing multiple junctions. This generalization will enables us to consider
the flux dependence of the transition rates in the two-junction split transmon and in the many-junction
fluxonium.
These two qubits are particular examples of the general case in which $M+1$ junctions separate $M+1$
superconducting islands forming
a loop. We use the convention that junction $j=0,\ldots,M$ is between
islands $j$ and $j+1$ and identify island $j=M+1$ with island $j=0$ --
see Fig.~\ref{fig:sp_tr}.
When the loop inductive energy is much larger than the Josephson energies of the
junctions (i.e., the loop inductance is small),
the phases are subject to the flux quantization constraint
\be\label{flux_quant}
\sum_{j=0}^M \varphi_j = 2\pi \Phi_e/\Phi_0 \, .
\ee
This constraint must be taken into account to derive the Hamiltonian $\hat H_{\{\phi\}}$ of the
$M$ independent phase degrees of freedom $\phi$, $\phi_k$ ($k=1,\ldots,M-1$)
starting from the Lagrangian\cite{ngnote} ${\cal L}_{\{\varphi\}}$ for the $M+1$ constrained phases $\varphi_j$
\be\label{lagra}
{\cal L}_{\{\varphi\}} = \sum_{j=0}^M \left[\frac{1}{2} C_j \left(\frac{\Phi_0}{2\pi} \dot{\varphi}_j\right)^2
+E_{Jj} \cos \varphi_j\right],
\ee
where the dot denotes derivative with respect to time, $C_j$ is the capacitance of junction $j$, and
$E_{Jj}$ its Josephson energy.
In Appendix~\ref{app:mj_ham} we derive the Hamiltonian assuming $M$ of the $M+1$ junctions
to be identical, which is relevant for both the split transmon ($M=1$)
and the fluxonium ($M\gg 1$). Explicit
expressions for the Hamiltonian in these two cases are presented below.

The total Hamiltonian $\hat H$ of the system consist of three terms, as in \eref{Htot}:
\be
\hat H = \hat H_{\{\phi\}} + \hat H_\qp + \hat H_T \, .
\ee
In addition to
$\hat H_{\{\phi\}}$ discussed above, the second contribution is the quasiparticle Hamiltonian
\be
\hat{H}_{\qp} = \sum_{j=0}^M \hat{H}_{\qp}^j \, , \quad \hat{H}_{\qp}^j = \sum_{n,\sigma} \epsilon_{n}^j
\hat\alpha^{j\dagger}_{n\sigma} \hat\alpha^j_{n\sigma}\, .
\ee
Here the index $j$ denotes the superconducting island; other symbols have the same meaning as in
\eref{Hqp} and we assume equal gaps in all islands, $\Delta^j \equiv \Delta$.
The final contribution to $\hat H$ is
the tunnel Hamiltonian, given by the following sum [cf. \eref{HTle}]
\be
\hat H_T = \sum_{j=0}^M \tilde{t}_j \!\!\sum_{n,m,\sigma}\!\!
i \sin \frac{\hat \varphi_j}{2}
\hat\alpha_{n\sigma}^{j\dagger} \hat\alpha^{j+1}_{m\sigma}
+ \text{H.c.}
\ee

The transition rate between qubit states can again be calculated using Fermi's golden rule as in \eref{FGR}.
We assume
that the quasiparticle distribution functions are the same in all islands and that tunneling across each
junction is not correlated with tunneling in nearby junctions -- this is a good assumption if the mean
level spacing in the finite size superconductors is small compared to the gap.
Then the total rate for the transition between
eigenstates of Hamiltonian $\hat{H}_{\{\phi\}}$ is
\be\label{wif_multi}
\Gif = \sum_{j=0}^M \left|\langle f_{\{\phi\}}|\sin \frac{\hat\varphi_j}{2}|i_{\{\phi\}}\rangle\right|^2
E_{Jj} \tilde{S}_{\qp}\left(\omega_{if}\right),
\ee
where for convenience we have extracted the Josephson energy prefactor from the spectral density,
$\tilde{S}_{\qp} = S_{\qp}/E_J$, with $S_{\qp}$ defined in \eref{tF_def}.
Similarly, the correction $\delta E_i$ to the energy of state $i_{\{\phi\}}$ is given by sums
over junctions which generalize \esref{de_EJ} and \rref{de_qp},
\bea\label{de_multi}
\delta E_i  & = & \delta E_{i,E_J} + \delta E_{i,\qp} \, ,\\
\delta E_{i,E_J} & = & \sum_{j=0}^{M} E_{Jj} \langle i_{\{\phi\}}|\cos \hat\varphi_j |i_{\{\phi\}} \rangle
(\ka + 2 \kaA) \, , \label{de_EJm} \\
\delta E_{i,\qp} & = & \sum_{j=0}^{M} E_{Jj} \!\sum_{k\neq i}
\left| \langle k_{\{\phi\}}|\sin \frac{\hat\varphi_j}{2} |i_{\{\phi\}} \rangle \right|^2\!
\tilde{F}_\qp(\omega_{ik}) , \quad \quad \ \label{de_qpm}
\eea
where $\tilde{F}_{\qp} = F_{\qp}/E_J$.
In the next subsections we use \eref{wif_multi} to calculate the transition rates for the
split transmon and the fluxonium, and \eref{de_multi} to find the frequency shift in the
split transmon. The flux-dependent transition rate between the two lowest even and odd states of a split Cooper
pair box has been recently considered in \ocite{schon} for gate voltage tuned at the operating point.

\subsection{Split transmon}
\label{sec:split_tr}

A split transmon consists of two junctions, $j=0,1$, in a superconducting loop, see Fig.~\ref{fig:sp_tr}.
Therefore,
there is only $M=1$ degree of freedom, which we denote with $\phi$, governed by the Hamiltonian
\be\label{tr_sp_ham0}
\hat H_\phi = 4E_C \hat N^2 - E_{J0} \cos (\hat\phi - 2\pi f)- E_{J1} \cos \hat\phi \, ,
\ee
see Appendix~\ref{app:mj_ham}. Here $\hat N = -id/d\phi$, $f=\Phi_e/\Phi_0$, and the charging energy
$E_C$ is related to the junctions' capacitances by
\be
E_C = \frac{e^2}{2(C_0+C_1)}\, .
\ee
Note that the Hamiltonian is periodic in $f$ with period 1, so we can assume $|f|\leq 1/2$ without loss of
generality (i.e., we can measure the normalized flux from the nearest integer).
After shifting $\phi \to \phi + \pi f$, the sum of the two Josephson terms can be rewritten as
\be
E_{J0} \cos (\hat\phi - 2\pi f) + E_{J1} \cos \hat\phi \to E_J(f) \cos (\hat \phi - \vartheta)\, ,
\ee
where the effective Josephson energy $E_J$ is modulated by the external flux
\be\label{EJ_flux}
E_J(f) = \left(E_{J0}+E_{J1}\right) \cos (\pi f) \sqrt{1+d^2\tan^2(\pi f)}
\ee
with
\be
d = \frac{E_{J0}-E_{J1}}{E_{J0}+E_{J1}}
\ee
and
\be\label{vt_def}
\tan \vartheta = d \tan (\pi f).
\ee
After a further shift $\phi \to \phi + \vartheta$ we arrive at
\be\label{tr_sp_ham}
\hat H_\phi = 4E_C \hat N^2 - E_{J}(f) \cos \hat\phi \, ,
\ee
which has the same form of the Hamiltonian for the single junction transmon [i.e., \eref{Hphi} with $E_L=0$]
but with a flux-dependent Josephson energy, \eref{EJ_flux}.
Therefore the spectrum follows directly from that of the single junction transmon (see Fig.~\ref{fig:trans})
and consists of nearly degenerate and well separated states.
The energy difference between well separated states is approximately given
by the flux-dependent frequency [cf. \eref{pl_fr}]
\be\label{opf_def}
\omega_p(f) = \sqrt{8E_C E_J(f)}.
\ee
Note that for the system to be in the transmon regime
\be\label{tr_reg}
E_J(f) \gg E_C
\ee
at some flux, a necessary condition is
\be
 E_{J0}+ E_{J1} \gg E_C \, .
\ee
Then we can distinguish two cases. First, in the nearly symmetric case of junctions with comparable
Josephson energies,  $|E_{J0}- E_{J1}| \lesssim E_C$, the condition \rref{tr_reg} is satisfied
not too close to half the flux quantum,
\be
|f|-1/2 \gg E_C/\pi(E_{J0}+ E_{J1}).
\ee
On the other hand, if the Josephson energies are sufficiently different, $|E_{J0}- E_{J1}| \gg E_C$, then
\eref{tr_reg} is satisfied at any flux.

The transition rate $\Gamma_{1 \to 0}$ between the
qubit states $|0\rangle$, $|1\rangle$
can be calculated using \eref{wif_multi} if we know the relation between $\varphi_j$
and $\phi$; the same relation is also needed to calculate the transition rate $\Gamma_{o \to e}$ between nearly
degenerate states -- see Appendix~\ref{app:eorate-sp} for details.
According to Appendix~\ref{app:mj_ham}, for the variable $\phi$ in \eref{tr_sp_ham0} we have
$\varphi_1 = \phi$ and $\varphi_0 = 2\pi f -\phi$. Accounting for the two changes of variables performed
to arrive at \eref{tr_sp_ham} we obtain
\be\label{sp_tr_var}\begin{split}
\hat\varphi_0 & = \pi f - \vartheta - \hat\phi \, , \\
\hat\varphi_1 & = \pi f + \vartheta + \hat\phi \, .
\end{split}\ee
In the transmon regime \rref{tr_reg}, we proceed as in the derivation of \eref{harm_mat_el} to find
\be\label{sp_tr_me0}
\left| \langle 0 | \sin \frac{\hat\varphi_j}{2} | 1 \rangle \right|^2 = \frac{E_C}{\omega_p(f)}
\frac{1+\cos(\pi f \pm \vartheta)}{2} \, ,
\ee
where the upper (lower) sign is to be used for $j=1$ ($j=0$).
Substituting this result
into \eref{wif_multi} and using \eref{Sqp_hf} with $\omega=\omega_p(f)$,
we find in the high frequency regime [cf. \eref{Gnn}]
\be
\Gamma_{1\to 0} =\frac{\ka}{2\pi} \sqrt{\frac{2\Delta}{\omega_p(f)}}
\frac{\omega_p^2(f)+\omega_p^2(0)}{\omega_p(f)} \, .
\ee

\begin{figure}
\includegraphics[width=0.46\textwidth]{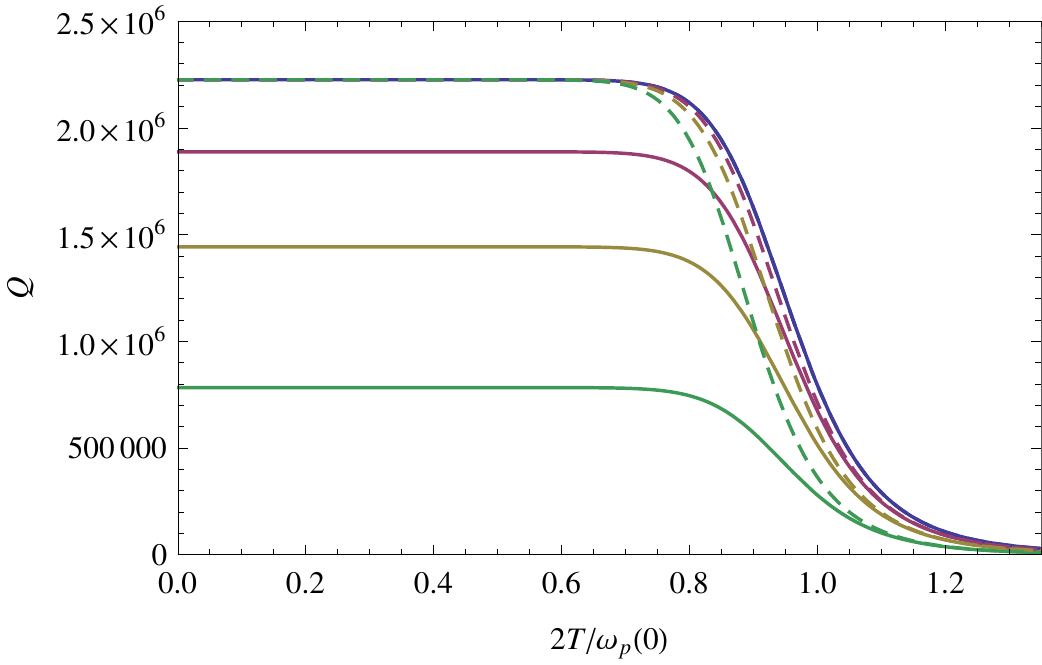}
\caption{Quality factor as function of $2T/\omega_p(0)$ in a symmetric split transmon. Solid lines are obtained
from \eref{Qqp_m} using a small nonequilibrium quasiparticle density $x_{ne} = 3.8\times 10^{-7}$ and a gap value
such that $\Delta/\omega_p(0) =6.9$ (these parameters are taken from experiments on single junction
transmon\cite{paik}). Flux increases from top to bottom; we show curves for $f=0$, $0.2$, $0.3$, and $0.4$,
respectively. We plot \eref{Qex_m} with dashed lines for the same values of flux. The extrinsic
quality factor is chosen so that solid and dashed lines match at $f=0$.}
\label{fig:Q_sp}
\end{figure}

For the transition quality factor we consider, as in Sec.~\ref{sec:qf_s}, the coexistence of equilibrium and
non-equilibrium quasiparticles [see \eref{f_eq_ne}] to find
\be\label{Qqp_m}\begin{split}
& \frac{1}{Q_{10}} = \frac{1}{2\pi}\left(1+\frac{\omega_p^2(0)}{\omega_p^2(f)}\right)
\bigg[
x_{ne}\sqrt{\frac{2\Delta}{\omega_p(f)}}  \\ & + 4
e^{-\Delta/T} \cosh\left(\frac{\omega_p(f)}{2 T}\right) K_0 \left(\frac{\omega_p(f)}{2 T}\right)
\bigg] .
\end{split}\ee
In Fig.~\ref{fig:Q_sp} we show with solid lines the quality factor as function of temperature for
4 different values of flux $f$ in a symmetric transmon (d=0).
As we discussed in Sec.~\ref{sec:qf_s}, an extrinsic relaxation mechanism could be limiting
the low temperature quality factor.
Characterizing this mechanism by a constant quality factor $Q_{\mathrm{ext}}$
and assuming that only equilibrium quasiparticles are present,
the transition quality factor has the form
\be\label{Qex_m}\begin{split}
\frac{1}{Q_{10,\mathrm{tot}}} = & \, \frac{1}{Q_{\mathrm{ext}}}
+ \frac{2}{\pi}
\left(1+\frac{\omega_p^2(0)}{\omega_p^2(f)}\right)  \\ & \times e^{-\Delta/T}
\cosh\left(\frac{\omega_p(f)}{2 T}\right) K_0 \left(\frac{\omega_p(f)}{2 T}\right).
\end{split}\ee
The dashed lines in Fig.~\ref{fig:Q_sp} show $Q_{10,\mathrm{tot}}$ as a function of temperature for the
same values of flux; the
quality factor $Q_{\mathrm{ext}}$
is chosen so that the zero-flux curve coincides
with the zero flux-curve described by \eref{Qqp_m}. The change of quality factor with flux is
markedly different in the two limiting cases (namely, presence of non-equilibrium
quasiparticles and no extrinsic relaxation mechanism vs. extrinsic relaxation with no non-equilibrium
quasiparticles) described by \esref{Qqp_m} and \rref{Qex_m}. Therefore the measurement of the temperature
and flux dependencies of the quality factor should give indications on the presence of non-equilibrium
quasiparticles. For example, the low-temperature measurements reported in \ocite{Houck} are compatible with a
flux-independent quality factor; to explain the data with \eref{Qqp_m} rather than \eref{Qex_m}
one would need to assume a
quasiparticle density that decreases with increasing flux. Since magnetic fields are known to
break pairs and thus increase the quasiparticle density, for the transmons considered in \ocite{Houck}
it is unlikely
that non-equilibrium quasiparticles are the source of the low-temperature qubit decay.

The frequency shift for the split transmon is obtained, as in Sec.~\ref{sec:freq_sh}, by calculating
the difference between correction to energies of nearby levels, \eref{dom}. The matrix
elements appearing in \esref{de_EJm}-\rref{de_qpm} are given by
\esref{harm_mat_el} and \rref{cos_me} with $\omega_{10}=\omega_p(f)$, $\varphi_0 = \theta + \pi f$ for $j=1$,
and $\varphi_0 = \theta - \pi f$ for $j=0$ [cf. \eref{sp_tr_var}-\rref{sp_tr_me0}]. Using
those expressions we find
\be
\delta E_{i+1,E_J} - \delta E_{i,E_J} = - \frac12 \omega_p(f) \left(\ka + 2\kaA \right)
\ee
and
\be\begin{split}
\delta E_{i+1,\qp} - \delta E_{i,\qp} = \frac{1}{16} \frac{\omega_p^2(0)+\omega_p^2(f)}{\omega_p(f)}
\\ \left[\tilde{F}_\qp (\omega_p(f)) + \tilde{F}_\qp (-\omega_p(f))\right].
\end{split}\ee
Then using the relation \rref{F_ImY} and \eref{Y2_hf}, we arrive at the high frequency result
\be\begin{split}
\frac{\delta \omega(f)}{\omega_p(f)} &\, =  \frac{1}{2}\Bigg\{
\kaA \left(\frac{\omega_p^2(0)}{\omega_p^2(f)}-1\right)
\\ & - \ka \left[\frac{1}{2\pi}\left(\frac{\omega_p^2(0)}{\omega_p^2(f)}+1\right)
\sqrt{\frac{2\Delta}{\omega_p(f)}}+1\right] \Bigg\}.
\end{split}\ee
At zero flux this expression agrees with \eref{do_hf} applied to a single junction transmon
($\EZO = \omega_p$, $\varphi_0 = 0$). However, similarly to the flux qubit, at finite
flux the split transmon
frequency shift is sensitive to the occupation of the Andreev bound states, see the first term in
curly brackets.

\subsection{Fluxonium}
\label{sec:fluxo}

In the fluxonium, an array of many identical junctions ($M \gg 1$) of Josephson energy
$E_{J1} \gg E_{C1}$ is connected to a weaker junction with $E_{J0} < E_{J1}$. Then the Hamiltonian
$\hat H_{\{\phi\}}$ for
the $M$ independent degrees of freedom can be approximately separated into independent terms for
the qubit phase $\phi$ and the $M-1$ phases $\phi_k$,
\bea\label{flux_ham}
\hat H_{\{\phi\}} & = &  \hat H_\phi + \sum_{k=1}^{M-1} \hat H_k \, , \\
\hat H_\phi & = & 4E_C\hat N^2 -E_{J0}\cos\hat\phi + \frac12 E_L(\hat \phi-2\pi f)^2, \nonumber \\
\hat H_k &=& 4E_{C1}\hat N_k^2 + \frac12 E_{J1} \hat\phi_k^2 \, , \nonumber
\eea
where
\be\label{fl_par}
E_L = \frac{E_{J1}}{M}, \qquad \frac{1}{E_C} = \frac{1}{E_{C0}} + \frac{1}{ME_{C1}},
\ee
see Appendix~\ref{app:mj_ham}.
There, we also give [\eref{nvi}] the relation between the $M+1$ (constrained) $\varphi_j$ variables and
the $M$ independent $\phi_k$ variables. Accounting for the
changes of variables that bring the Hamiltonian in the form given above, we have schematically
\be\begin{split}
\varphi_0 & = \phi \, ,\\
\varphi_j & = L_j(\{\phi_k\}) +(2\pi f - \phi)/M\, , \qquad j=1,\ldots, M \, .
\end{split}\ee
Here $L_j(\{\phi_k\})$ denote linear combinations of the variables $\phi_k$, $k=1,\ldots,M-1$, whose specific
form can be found in Appendix~\ref{app:mj_ham} but is not needed here, while we show explicitly the
dependence of the constrained variables $\varphi_j$ on the qubit phase $\phi$.
As in the previous section, we take $|f|<1/2$ without loss of generality.

As an example of the calculation of the transition rate for such a system, we assume that the
plasma frequency $\omega_{p1}=\sqrt{8E_{C1}E_{J1}}$ of the array junctions is larger than the other
relevant energy scales (namely, quasiparticle energy $\delta E$ and qubit frequency $\EZO$).
Then we can take the many-body
state of the system $|\Psi_{\{\phi\}}\rangle$ in the product form
\be\label{psi_fl}
|\Psi_{\{\phi\}}\rangle = |\psi_\phi\rangle \prod_{k=1}^{M-1} |0_k\rangle,
\ee
where $|\psi_\phi\rangle$ is a low-energy eigenstate of $\hat H_\phi$ and $|0_k\rangle$
is the ground state wave function of the $k$th oscillator. The approximations used to derive
$\hat H_{\{\phi\}}$ in \eref{flux_ham} imply that in the formula \rref{wif_multi} for the transition
rate we can linearize the sine for $j=1,\ldots,M$. Therefore,
for the transition rate between two
states of the form \rref{psi_fl} we obtain
\be\label{Gif_fl1}\begin{split}
\Gif = \tilde{S}_{\qp}\left(\omega_{if}\right)
\Bigg[E_{J0}\left|\langle f_\phi|\sin \frac{\hat\phi}{2}|i_\phi\rangle\right|^2
\\
 + E_{L} \left|\langle f_\phi|\frac{\hat\phi}{2}|i_\phi\rangle\right|^2\Bigg].
\end{split}\ee

In the weak tunneling limit $\bar\epsilon \ll 2\pi^2 E_L \ll \omega_p = \sqrt{8E_CE_{J0}}$
(with $\bar\epsilon$ the tunnel splitting of the qubit states at $f=1/2$),
we can use
directly the results of Sec.~\ref{sec:fl_qb}: the flux-dependent qubit frequency $\EZO(f)$
is given by \eref{o_f} and the first excited state $|i_\phi\rangle = |+\rangle$ and ground state
$|f_\phi\rangle = |-\rangle$ are the linear combination of states localized in wells $m=0$, $1$
in \eref{f_s}.
For the first term in square brackets in \eref{Gif_fl1}, the matrix element
is given by \eref{fl_qb_me}.
To evaluate the second term in the same regime, we note that for states $|m\rangle$,
$|j\rangle$ -- that is, states localized in wells $m$ and $j$ as
in \eref{loc_st} -- we have
\be
\langle j|\frac{\hat\phi}{2}|m\rangle = \pi\left[m\left(1-\frac{E_L}{E_{J0}}\right)+
\frac{E_L}{E_{J0}} f \right]\delta_{j,m} \, .
\ee
Therefore for the states in \eref{f_s} we find
\be\label{lin_me}
\left|\langle -|\frac{\hat\phi}{2}| + \rangle\right|^2 =
\left(\frac{\pi}{2}\right)^2\left(1-\frac{E_L}{E_{J0}}\right)^2 \frac{\bar\epsilon^2}{\EZO^2(f)} \, .
\ee
In contrast to the matrix element of $\sin\hat\varphi/2$ considered in Sec.~\ref{sec:fl_qb},
the contribution due to tunneling can be neglected in this case.
Substituting this result and the leading term from \eref{fl_qb_me} into the square brackets of
\eref{Gif_fl1} we get\cite{ren_neg}
\be\label{flum_me}\begin{split}
\frac{\pi^2}{4} \frac{\bar\epsilon^2}{\EZO^2(f)} E_L
\Bigg[ (2\pi)^2\frac{E_L}{E_{J0}} \left(f-\frac12\right)^2
\left(\sqrt{2}D\right)^2 \left(\frac{E_{J0}}{E_C}\right)^{2/3}
\\ + \left(1-\frac{E_L}{E_{J0}}\right)^2 \Bigg].
\end{split}\ee
In this expression, the first term in square brackets originates from the weak junction and the
second one from the array.
Note that when considering
flux near half the flux quantum
we can neglect the first term in comparison to the second
and the losses due to the array dominate over those due to the weak junction.
Keeping only the leading contribution in \eref{flum_me}
and using \eref{Sqp_hf} in \eref{Gif_fl1}, we arrive at
the expression for the rate in the high-frequency regime
\be
\Gamma_{+\to -} = \ka \sqrt{\frac{2\Delta}{\EZO(f)}}\, 2\pi E_L \frac{\EZO^2(1/2)}{\EZO^2(f)}
\ee
with $\EZO(f)$ defined in \eref{o_f}. Note that since the frequency increases as the
reduced flux $f$ moves away from $1/2$, the transition rate is the largest at half the flux quantum.

\section{Summary}
\label{sec:summ}

In this work we have presented in detail a general approach to study the effects of quasiparticles
on relaxation and frequency of
superconducting qubits. The theory is applicable to any qubit -- the case of single-junction
systems is considered in Sec.~\ref{sec:th_s} and the generalization to multi-junction ones is given
in Sec.~\ref{sec:multi}. Our analysis is valid for both thermal equilibrium quaisparticles
and arbitrary non-equilibrium distributions, so long as the quasiparticle energy is small compared
to the qubit frequency -- this condition, not necessary in thermal equilibrium,
 ensures that quasiparticles primarily cause relaxation and not excitation of the qubit.

For single-junction qubits, we have studied in Sec.~\ref{sec:semi} the weakly anharmonic
limit. For small phase fluctuations, both quality factor (Sec.~\ref{sec:qf_s}) and frequency
shift (Sec.~\ref{sec:freq_sh}) are determined by
transitions between neighboring qubit levels and can be related to real and imaginary
part of the ``classical'' junction admittance, respectively.  The small fluctuation case applies to
phase and transmon qubits and our results in \esref{Q1_nt} and \rref{do_hf} have been
successfully tested in recent experiments~\cite{paik,martinis2} with these qubits.
For strong anharmonicity, we have presented in Sec.~\ref{sec:further} results for the quasiparticle
transition rate in the Cooper pair box and the flux qubit.

We have considered two examples of multi-junction qubits, the two-junction split transmon in
Sec.~\ref{sec:split_tr} and the many-junction fluxonium in Sec.~\ref{sec:fluxo}. In particular,
we argue that measuring the temperature and flux dependencies of the quality factor of a
split transmon could help resolve the question
of wether non-equilibrium quasiparticles are present
at low temperatures, see \esref{Qqp_m}-\rref{Qex_m} and Fig.~\ref{fig:Q_sp}.

\acknowledgments

We thank L. Frunzio, A. Kamal, and J. Koch for stimulating discussions and help with numerical calculations.
This research was funded by
DOE (Contract DE-FG02-08ER46482), by Yale University, and by
the Office of the Director of National
Intelligence (ODNI), Intelligence Advanced Research Projects
Activity (IARPA), through the Army Research Office (Contract No. W911NF-09-1-0369).  All statements
of fact, opinion or conclusions contained herein are those of the
authors and should not be construed as representing the official
views or policies of IARPA, the ODNI, or the U.S. Government.

\appendix

\section{Correction to energy levels}
\label{app:encorr}

To calculate the correction to the energy levels as presented in Sec.~\ref{sec:encorr}, we must account for
both quasiparticle and pair tunneling. Note that due to energy conservation
the latter does not affect the transition rate
$\Gamma_{i\to f}$ between states $|i\rangle$ and $|f\rangle$ so long as $\omega_{if} < 2\Delta$;
for this reason the pair tunneling Hamiltonian $H_T^p$ was neglected in \eref{Htot}.
More generally, the total Hamiltonian of the single junction system is
\be\label{Htot_p}
\hat{H}_\mathrm{tot} = \hat{H}_0 +\hat{H}_T + \hat{H}_T^p  + \hat{H}_{E_J}
\ee
with
\be\label{H0}
\hat{H}_0 = \hat{H}_\varphi + \hat{H}_\qp \, .
\ee
The Hamiltonians $\hat{H}_\varphi$, $\hat{H}_\qp$, and $\hat{H}_T$ are defined in
\esref{Hphi}, \rref{Hqp}, and \rref{HT}, respectively, and the pair tunneling term is
\be\label{HTp}\begin{split}
& \hat{H}_T^p = \tilde{t} \sum_{n,m}\bigg[\left(e^{i\frac{\hat\varphi}{2}} u_n^L v_m^R +
e^{-i\frac{\hat\varphi}{2}} u_m^R v_n^L \right)\hat\alpha_{n\uparrow}^{L\dagger}
\hat\alpha^{R\dagger}_{m\downarrow} \\ & +\left(e^{-i\frac{\hat\varphi}{2}} v_m^R u_n^L +
e^{i\frac{\hat\varphi}{2}}
v_n^L u_m^R \right)\hat\alpha_{m\downarrow}^{R} \hat\alpha^L_{n\uparrow} \bigg] + (L \leftrightarrow
R).
\end{split}\ee
The last term in \eref{Htot_p},
\be
\hat{H}_{E_J} = E_J \cos \hat\varphi \, ,
\ee
is necessary to avoid ``double counting'':
the Josephson energy originates from pair tunneling,
so its inclusion in the effective Hamiltonian $\hat{H}_\varphi$, \eref{Hphi}, must be compensated
for by subtracting the same term here.
We will show below that this treatment is justified for small quasiparticle density.

In both the quasiparticle tunneling Hamiltonian $\hat{H}_T$, \eref{HT}, and the
pair tunneling one in \eref{HTp}, using the definitions given after \eref{Hqp}
the (real) Bogoliubov amplitudes are
\be
\left(u^j_n\right)^2 = 1-\left(v^j_n\right)^2 = \frac12\left(1+\frac{\xi^j_n}{\epsilon_n^j}\right),
\qquad j=L,R \, .
\ee
As in the main text, we assume equal gaps and distribution functions in the leads,
$\Delta^L=\Delta^R\equiv \Delta$ and $f^L = f^R \equiv f$. Moreover, we neglect the
contributions of the charge mode $f_Q(\epsilon) = (f(\xi)-f(-\xi))/2$, since they are suppressed by
the small factor $\delta E/\Delta\ll 1$ compared to the leading contributions due to the energy mode
$f_E$, \eref{fE}; for simplicity, in this Appendix we drop the subscript $E$.

We want to evaluate the correction $\delta E_i$ to the energy of level $i$ of the qubit
at second order in the tunneling amplitude $\tilde{t}$ for small quasiparticle density. Thus,
$\hat{H}_0$ in \eref{H0} is the unperturbed Hamiltonian, and we distinguish
three contributions to $\delta E_i$,
\be\label{dE_i_app}
\delta E_i = \delta E_i^{(1)}+ \delta E_i^{(2)} + \delta E_i^{(3)},
\ee
caused respectively by $\hat{H}_T$, $\hat{H}_T^p$, and $\hat{H}_{E_J}$.
Noting that the latter is already
of second order in $\tilde{t}$, we treat it within first order perturbation theory to write
\be\label{de_i3}
\delta E_i^{(3)} = E_J \langle i | \cos\hat\varphi| i \rangle \, .
\ee

The quasiparticle tunneling correction $\delta E_i^{(1)}$ is obtained by second order perturbation theory,
\be\label{dE_i1_pt}
\delta E_i^{(1)} =  - \sum_{k,\{\lambda\}_\qp} \langle\!\langle
\frac{|\langle k,\{\lambda\}_\qp|\hat{H}_T|i,\{\eta\}_\qp \rangle |^2}
{E_{\lambda,\qp} - E_{\eta,\qp} - \omega_{ik}} \rangle\!\rangle_\qp\, ,
\ee
where
\be
\omega_{ik} = E_k - E_i
\ee
and the notation is the same as in Sec.~\ref{sec:th_s}: $\{\eta\}_\qp$ and $\{\lambda\}_\qp$ denote
quasiparticle states, $E_{\lambda,\qp}$ and $E_{\eta,\qp}$ their energies, and
$\langle\!\langle \ldots \rangle\!\rangle_\qp$ averaging over $\{\eta\}_\qp$. Performing the averaging,
after lengthy but straightforward algebra we arrive at
\bea\label{dE_i1}
&& \delta E_i^{(1)} =\frac{4E_J}{\pi^2 \Delta} P \sum_k \int_{\Delta_\qp}^\infty\!\!d\epsilon_L
\int_{\Delta_\qp}^\infty\!\!d\epsilon_R \\
&& \left[
\left|\langle k | \sin\frac{\hat\varphi}{2}|i\rangle \right|^2
\!A_+(\epsilon_L,\epsilon_R)
+ \left|\langle k | \cos\frac{\hat\varphi}{2}|i\rangle \right|^2
\!A_-(\epsilon_L,\epsilon_R) \right] \nonumber \\
&& \qquad \quad \quad \times \left[\frac{f(\epsilon_L)(1-f(\epsilon_R))}{\epsilon_L-\epsilon_R -\omega_{ik}} -
\frac{(1-f(\epsilon_L))f(\epsilon_R)}{\epsilon_L-\epsilon_R +\omega_{ik}}\right], \nonumber
\eea
where we introduced the functions
\be\begin{split}
A_\pm (\epsilon_L,\epsilon_R) & =
\frac{\epsilon_L}{\sqrt{\epsilon_L^2-\Delta_\qp^2}}
\frac{\epsilon_R}{\sqrt{\epsilon_R^2-\Delta_\qp^2}} \\ & \pm
\frac{\Delta_\qp}{\sqrt{\epsilon_L^2-\Delta_\qp^2}}
\frac{\Delta_\qp}{\sqrt{\epsilon_R^2-\Delta_\qp^2}}
\end{split}\ee
describing combinations of BCS densities of states. Both these functions and the lower
integration limit depend on the self-consistent gap $\Delta_\qp$; however, since the integrand
in \eref{dE_i1} is at least linear in the distribution function $f$, we can neglect the
gap suppression by quasiparticles [see \eref{delqp}] and approximate
$\Delta_\qp \simeq \Delta$.

We note that the combinations of distribution functions
in the last term of \eref{dE_i1} restricts to low energies only one of the energy integrals, while
the other integral is logarithmically divergent. To isolate this divergence, we add and subtract the
term obtained by setting $\omega_{ik}=0$ in the denominator; more precisely, we define
\be\label{p_diff}
P\frac{1}{\epsilon_L - \epsilon_R} = \frac12 \lim_{\omega \to 0^+}
\frac{1}{\epsilon_L - \omega - \epsilon_R} + \frac{1}{\epsilon_L + \omega - \epsilon_R}
\ee
and separate in $\delta E_i^{(1)} = \delta E_i^{(1),f}+\delta E_i^{(1),d}$ a finite term,
\bea\label{dE_i1f}
&& \delta E_i^{(1),f} = \frac{8E_J}{\pi^2 \Delta} P \sum_{k\neq i} \int_{\Delta}^\infty\!\!d\epsilon_L
\int_{\Delta}^\infty\!\!d\epsilon_R \\
&& \left[
\left|\langle k | \sin\frac{\hat\varphi}{2}|i\rangle \right|^2
\!A_+(\epsilon_L,\epsilon_R)
+ \left|\langle k | \cos\frac{\hat\varphi}{2}|i\rangle \right|^2
\!A_-(\epsilon_L,\epsilon_R) \right] \nonumber \\
&& \quad \qquad \times f(\epsilon_L)(1-f(\epsilon_R))\left[\frac{1}{\epsilon_L-\epsilon_R -\omega_{ik}} -
\frac{1}{\epsilon_L-\epsilon_R}\right], \nonumber
\eea
from a divergent one,
\be\label{dE_i1d}\begin{split}
\delta E_i^{(1),d} = & \frac{4E_J}{\pi^2 \Delta} P\!\int_{\Delta}^\infty\!\!d\epsilon_L
\int_{\Delta}^\infty\!\!d\epsilon_R \, \frac{f(\epsilon_L)-f(\epsilon_R)}{\epsilon_L-\epsilon_R}
\\ & \Bigg[\frac{\epsilon_L}{\sqrt{\epsilon_L^2-\Delta^2}}
\frac{\epsilon_R}{\sqrt{\epsilon_R^2-\Delta^2}}  \\ & - \langle i | \cos \hat\varphi |i\rangle
\frac{\Delta}{\sqrt{\epsilon_L^2-\Delta^2}}
\frac{\Delta}{\sqrt{\epsilon_R^2-\Delta^2}}\Bigg].
\end{split}\ee
To obtain this expression we used the identities
\be\label{skp}
\sum_k \left| \langle k | \sin\frac{\hat\varphi}{2}|i\rangle \right|^2+
\left| \langle k | \cos\frac{\hat\varphi}{2}|i\rangle \right|^2 = 1
\ee
and
\be\label{skm}
\sum_k \left| \langle k | \sin\frac{\hat\varphi}{2}|i\rangle \right|^2 -
\left| \langle k | \cos\frac{\hat\varphi}{2}|i\rangle \right|^2 = - \langle i | \cos\hat\varphi|i\rangle \, .
\ee

Equation \rref{dE_i1f} for $\delta E_i^{(1),f}$ can be further simplified using the relations
\bea\label{Aapp}
&& \epsilon_L \epsilon_R \left[\frac{1}{\epsilon_L-\epsilon_R -\omega_{ik}} -
\frac{1}{\epsilon_L-\epsilon_R }\right] \\ && = \epsilon_L^2
\left[\frac{1}{\epsilon_L-\epsilon_R -\omega_{ik}} -
\frac{1}{\epsilon_L-\epsilon_R }\right] - \frac{\epsilon_L \omega_{ik}}{\epsilon_L-\epsilon_R -\omega_{ik}}
\nonumber \\ && \simeq \Delta^2 \left\{
\left[\frac{1}{\epsilon_L-\epsilon_R -\omega_{ik}} -
\frac{1}{\epsilon_L-\epsilon_R }\right]
- \frac{\omega_{ik}/\Delta}{\epsilon_L-\epsilon_R -\omega_{ik}} \right\}, \nonumber
\eea
where the approximation is valid because the distribution function restricts the integral over
$\epsilon_L$ to low energies above the gap. As discussed in Sec.~\ref{sec:th_s}, the matrix
elements of operators $e^{\pm i\hat\varphi/2}$ describe the transfer of a single charge.
For this reason, for a low-lying level $i$ the main contribution in the sum over
states $k$ comes either from levels with energy difference $\omega_{ik} \sim E_C \ll \Delta$ (when $E_C$ is
large compared to $E_L$, $E_J$), or from nearby levels (for small $E_C$). In both cases we have
$\omega_{ik} \ll \Delta$, since at large energy differences the matrix elements quickly decrease;
this is evident, for example, in the expressions for the matrix elements in Sec.~\ref{sec:semi}.
Then according to \eref{Aapp} the term proportional to $A_-$ is suppressed by the small parameter
$\omega_{ik}/\Delta$ in comparison to the leading term in $A_+$, and we can approximate
$\delta E_i^{(1),f}$ as
\bea\label{dE_i1ff}
&& \delta E_i^{(1),f} \simeq \frac{16E_J}{\pi^2 \Delta} P \sum_{k\neq i} \int_{\Delta}^\infty\!\!d\epsilon_L
\int_{\Delta}^\infty\!\!d\epsilon_R \\
&& \left|\langle k | \sin\frac{\hat\varphi}{2}|i\rangle \right|^2
\frac{\Delta}{\sqrt{\epsilon_L^2-\Delta^2}}
\frac{\Delta}{\sqrt{\epsilon_R^2-\Delta^2}} \nonumber \\ &&
\times f(\epsilon_L)(1-f(\epsilon_R))\left[\frac{1}{\epsilon_L-\epsilon_R -\omega_{ik}} -
\frac{1}{\epsilon_L-\epsilon_R}\right]. \nonumber
\eea
Defining the function $F_\qp$ by
\be\label{Fqp_def}\begin{split}
F_{\qp}(\omega) = \frac{16 E_J}{\pi^2 \Delta} P\!\int_\Delta^\infty\!\!d\epsilon_L
\int_\Delta^\infty\!\!d\epsilon_R
\frac{\Delta}{\sqrt{\epsilon_L^2-\Delta^2}} \frac{\Delta}{\sqrt{\epsilon_R^2-\Delta^2}} \\
f(\epsilon_L)(1-f(\epsilon_R))\left[\frac{1}{\epsilon_L-\epsilon_R -\omega} -
\frac{1}{\epsilon_L-\epsilon_R }\right]
\end{split}\ee
we arrive at the expression for the quasiparticle correction to the energy $\delta E_{i,\qp}$
given in \eref{de_qp}.

The treatment of the pair correction term $\delta E_i^{(2)}$ in \eref{dE_i_app} is similar to the
above one for $\delta E_i^{(1)}$. The pair correction is found by calculating the matrix element of
$\hat{H}_T^p$ [\eref{HTp}] rather than $\hat{H}_T$ in \eref{dE_i1_pt}; we find
\bea
&& \delta E_i^{(2)} =\frac{4E_J}{\pi^2 \Delta} P \sum_k \int_{\Delta_\qp}^\infty\!\!d\epsilon_L
\int_{\Delta_\qp}^\infty\!\!d\epsilon_R \\
&& \left[
\left|\langle k | \sin\frac{\hat\varphi}{2}|i\rangle \right|^2
\!A_-(\epsilon_L,\epsilon_R)
+ \left|\langle k | \cos\frac{\hat\varphi}{2}|i\rangle \right|^2
\!A_+(\epsilon_L,\epsilon_R) \right] \nonumber \\
&& \quad \quad \quad \times \left[\frac{f(\epsilon_L)f(\epsilon_R)}{\epsilon_L+\epsilon_R -\omega_{ik}} -
\frac{(1-f(\epsilon_L))(1-f(\epsilon_R))}{\epsilon_L+\epsilon_R +\omega_{ik}}\right]. \nonumber
\eea
Note that in this expression there is a term independent of the distribution function, for which
the approximation $\Delta_\qp \simeq \Delta$ is not applicable.
Since $\epsilon_L+\epsilon_R\ge 2\Delta_\qp$, repeating the argument preceding \eref{dE_i1ff}
we can neglect $\omega_{ik}$ in the denominator and use identities \rref{skp}-\rref{skm} to obtain
\bea
\delta E_i^{(2)} & \simeq & \frac{4E_J}{\pi^2 \Delta} P\!\int_{\Delta_\qp}^\infty\!\!d\epsilon_L
\int_{\Delta_\qp}^\infty\!\!d\epsilon_R \, \frac{f(\epsilon_L)+f(\epsilon_R)-1}{\epsilon_L+\epsilon_R}
\nonumber \\ && \Bigg[\frac{\epsilon_L}{\sqrt{\epsilon_L^2-\Delta_\qp^2}}
\frac{\epsilon_R}{\sqrt{\epsilon_R^2-\Delta_\qp^2}} \\ && + \langle i | \cos \hat\varphi |i\rangle
\frac{\Delta_\qp}{\sqrt{\epsilon_L^2-\Delta_\qp^2}}
\frac{\Delta_\qp}{\sqrt{\epsilon_R^2-\Delta_\qp^2}}\Bigg]. \nonumber
\eea
Both in this expression and in \eref{dE_i1d}, the first term in square bracket does not depend
on the level index $i$. Therefore, it leads to an unimportant common shift of all the levels which
we neglect.~\cite{renorm} Keeping only the second term in each square brackets, we write
\be
\delta E_i^{(1),d} + \delta E_i^{(2)} \approx \delta E_i^\Delta + \delta E_i^A,
\ee
where, separating the terms independent of and proportional to the distribution function $f$, we have
\be\begin{split}
\delta E_i^\Delta = -\frac{4E_J}{\pi^2 \Delta}  \langle i | \cos \hat\varphi |i\rangle
P\!\int_{\Delta_\qp}^\infty\!\!d\epsilon_L \int_{\Delta_\qp}^\infty\!\!d\epsilon_R \\
\frac{\Delta_\qp}{\sqrt{\epsilon_L^2-\Delta_\qp^2}}\frac{\Delta_\qp}{\sqrt{\epsilon_R^2-\Delta_\qp^2}}
\frac{1}{\epsilon_L+\epsilon_R}
\end{split}\ee
and
\be\label{de_iA}\begin{split}
\delta E_i^A = \frac{8E_J}{\pi^2 \Delta} \langle i | \cos \hat\varphi |i\rangle
\, P\!\int_{\Delta}^\infty\!\!d\epsilon_L  \frac{\Delta}{\sqrt{\epsilon_L^2-\Delta^2}} f(\epsilon_L) \\
\times \int_{\Delta}^\infty\!\!d\epsilon_R \frac{\Delta}{\sqrt{\epsilon_R^2-\Delta^2}}
\left[\frac{1}{\epsilon_L+\epsilon_R} - \frac{1}{\epsilon_L-\epsilon_R}\right].
\end{split}\ee
In both expressions the integrations can be performed analytically [using in \eref{de_iA} the
definition \rref{p_diff}]. We obtain
\be
\delta E_i^\Delta = - \frac{E_J}{\Delta} \Delta_\qp \langle i | \cos \hat\varphi |i\rangle
\ee
and
\be
\delta E_i^A = 2 E_J f(\Delta) \langle i | \cos \hat\varphi |i\rangle \, .
\ee
Finally, using \esref{alp_def}, \rref{delqp}, and \rref{de_i3} we arrive at
\be
\delta E_i^\Delta + \delta E_i^A + \delta E_i^{(3)} = E_J \left(\ka + 2 \kaA\right)
\langle i | \cos \hat\varphi |i\rangle \, ,
\ee
which is the correction $\delta E_{i,E_J}$ in \eref{de_EJ}. This result, together with
\esref{dE_i1ff}-\rref{Fqp_def}, concludes the derivation of the
formulas presented in Sec.~\ref{sec:encorr}.

\section{Gate-dependent energy splitting in the transmon}
\label{app:eosplit}

The transmon low-energy spectrum is characterized by well separated [by the plasma frequency
$\omega_p$, \eref{pl_fr}] and nearly degenerate levels whose energies, as shown in
Fig.~\ref{fig:trans},
vary periodically with the gate voltage $n_g$.
Here we derive the asymptotic expression (valid
at large $E_J/E_C$) for the energy splitting between the nearly degenerate levels. We consider first the
two lowest energy states and then generalize the result to higher energies.

Using the notation of Sec.~\ref{sec:th_s}, the transmon Hamiltonian is
\be\label{Hphi_tr}
\hat{H}_{\varphi} = 4E_C \left(\hat{N}-n_g\right)^2 -E_J \left(1+\cos \hat\varphi\right) \, .
\ee
Its eigengstates can be written exactly in terms of Mathieu functions.~\cite{transmon}
However, since $E_J \gg E_C$
a tight-binding approach~\cite{harrison} can be used in which the two lowest (even and odd) eigenstates
$\Psi^{e}$ and $\Psi^o$ are given by sums of localized wavefunctions,
\bea\label{Psieo}
\Psi^{e} (\varphi ; n_g) & = & e^{in_g \varphi} \frac{1}{\sqrt{L}}\sum_j \psi(\varphi - 2\pi j)
e^{-i n_g 2\pi j}, \\
\Psi^{o} (\varphi ; n_g) & = & e^{in_g \varphi} \frac{1}{\sqrt{L}}\sum_j \psi(\varphi - 2\pi j)
e^{-i n_g 2\pi j}e^{-i\pi j}, \nonumber
\eea
where $L\gg 1$ is the number of sites, labeled with index $j$, and $\psi$ is the ground state of the Hamiltonian
\be\label{H_w}
\hat{H} = 4E_C \hat{N}^2 + V(\hat\varphi)
\ee
with
\be\label{atpot}
V(\varphi) = \left\{
\begin{array}{ll}
-E_J(1+\cos\varphi), & |\varphi| < \pi \\
0, & |\varphi| > \pi
\end{array}\right. \, .
\ee
This potential is such that $\sum_j V(\varphi-2\pi j) = -E_J(1+\cos\varphi)$.
Note that the even (odd) state is a linear combination of even (odd) charge eigenstates, as can be shown
by considering the overlap of $\Psi^{e(o)}$ with the charge eigenstate $e^{i n \varphi/2}$
for arbitrary integer $n$ [in \eref{Psieo} what distinguish the odd state from the even one is the last
exponential in the expression for $\Psi^o$, which changes the sign of the localized
wavefunction at odd sites $j$].

The energy difference $\omega_{eo}$ between the two states is
\be
\omega_{eo} = \langle \Psi^o| \hat{H}_\varphi |\Psi^o\rangle
-  \langle \Psi^e| \hat{H}_\varphi |\Psi^e\rangle \, .
\ee
Using \eref{Psieo}, the contributions to $\omega_{eo}$ due to products of wavefunctions $\psi$ localized at the
same site cancel. The leading contribution to $\omega_{eo}$ originates from products of wavefunctions localized
at nearby sites,
\be\label{de_ng}
\omega_{eo} = \epsilon_0 \cos \left(2\pi n_g\right)
\ee
with
\be\label{te0}
\epsilon_0 = -4 \int d\varphi \, \psi(\varphi) \psi(\varphi -2\pi) V(\varphi) \, .
\ee
To estimate the above integral, the behavior of the wavefunction $\psi$ near $\varphi = \pi$
is needed; in this region a good approximation is given by the semiclassical wavefunction~\cite{landau}
\be\label{qcwf}
\psi (\varphi) \simeq \left\{ \begin{array}{r}
\frac{C_0}{2\sqrt{p(\varphi)}}
\exp\left[ -\int_a^\varphi \!d\phi\, p(\phi) \right],
\qquad \ a < \varphi < \pi \\
A_0 \exp \left[ -\sqrt{\frac{E_J}{2E_C}\left(1-\frac{\omega_p}{4E_C}\right)} (\varphi - \pi )\right],
  \varphi > \pi \end{array}\right.
\ee
where $C_0$ and $A_0$ are constants,
\be\label{p_def}
p(\varphi) = \sqrt{\frac{E_J}{4E_C}} \sqrt{1-\frac{\omega_p}{2E_J} - \cos\varphi} \ ,
\ee
and $a$ is the classical turning point defined by $p(a)=0$.
The constant $C_0$ is determined by the normalization condition of the wavefunction, and
$A_0$ then follows from continuity of the wavefunction. For states
with large quantum number the semiclassical approximation can be used also in the classically
accessible region $|\varphi| < a$; the corresponding estimate for the normalization constant, which
we indicate with $C_\infty$, is $C_\infty = \sqrt{\omega_p/4E_C \pi}$ -- see \ocite{landau}.
Here we are interested in the ground state (and more generally in low-lying states), for which
$C_\infty$ is known to underestimate the normalization factor.\cite{jcp,furry} To evaluate
$C_0$ we note that for $|\varphi| \ll \pi$ the potential $V(\varphi)$ in \eref{atpot} is well approximated
by that of the harmonic oscillator; therefore the semiclassical wavefunction \rref{qcwf}
should match the normalized wavefunction of the harmonic oscillator
given in \eref{loc_st} (with $\varphi_m =0$) in the region $a\ll \varphi \ll \pi$.
Indeed, in this region we expand
the cosine in \eref{p_def} and rescale variables ($\phi = \tilde{\phi}\sqrt{\omega_p/E_J}$) to find
\bea
\int_a^\varphi \!d\phi\, p(\phi) & \simeq & \int_1^{\tilde\varphi}\!d\tilde{\phi}
\sqrt{\tilde{\phi}^2-1} \\
& =  &\frac{1}{2}\left[\tilde\varphi\sqrt{\tilde{\varphi}^2-1} -
\ln \left(\tilde\varphi+\sqrt{\tilde{\varphi}^2-1}\right)
\right] \nonumber
\\ & \simeq & \frac{1}{2}\frac{E_J}{\omega_p}\varphi^2 - \frac{1}{4} - \frac12
\ln\left(2\varphi\sqrt{E_J/\omega_p}\right).\nonumber
\eea
Using this expression, and $p(\varphi)\simeq \varphi \sqrt{E_J/8E_C}$ in the denominator,
\eref{qcwf} becomes
\be
\psi(\varphi) \simeq  C_0\frac{e^{1/4}}{\sqrt{2}}\left(\frac{8E_C}{E_J}\right)^{1/8}
e^{-\varphi^2E_J/2\omega_p}.
\ee
This function matches \eref{loc_st} by setting
\be
C_0 = \sqrt{\frac{\omega_p}{4E_C}} (\pi e)^{-1/4} = C_\infty \left(\frac{\pi}{e}\right)^{1/4}.
\ee
The last form shows that the correct normalization factor is larger than the usual semiclassical
estimate.

Having found the normalization constant, we now consider the wavefunction in the region
near $\varphi = \pi$. There we can further simplify \eref{qcwf} as follows: we rewrite
the integral in the exponential in the first line of \eref{qcwf} as
\be\label{intsplit}
\int_a^\varphi \!d\phi\, p(\phi) = \int_a^\pi \!d\phi\, p(\phi) - \int_\varphi^\pi \!d\phi\, p(\phi) \,.
\ee
Then the first integral on the right hand side is
\be
\int_a^\pi \!d\phi\, p(\phi) = \sqrt{\frac{2E_J}{E_C}} \left[E(k) - (1-k^2) K(k) \right],
\ee
where $E$ and $K$ denote the complete elliptic integrals with modulus $k$, which has the value
\be\label{kdef}
k^2 \equiv 1- k'^2 = 1- \frac{\omega_p}{4E_J} \, .
\ee
Here we are interested in the limit $k\to 1$, in which the complete elliptic integrals behave as
\be\begin{split}
E(k) & \simeq 1 + \frac{1}{2} k'^2 \left(\ln \frac{4}{k'} -\frac 12 \right) \, ,\\
K(k) & \simeq \ln\frac{4}{k'} \, .
\end{split}\ee
The last integral in \eref{intsplit} can be approximated as
\be\label{int2app}\begin{split}
\int_\varphi^\pi \!d\phi\, p(\phi) \simeq \sqrt{\frac{E_J}{2E_C}}
\bigg[\sqrt{1-\frac{\omega_p}{4E_J}}(\pi-\varphi)
\\ -\frac{(\pi-\varphi)^3}{24\sqrt{1-\omega_p/4E_J}}\bigg].
\end{split}\ee
Substituting \esref{intsplit}-\rref{int2app} into \eref{qcwf},
using $p(\pi) \simeq \sqrt{E_J/2E_C}$ in the square root in the denominator of the first line,
and requiring continuity of the wavefunction, we arrive at
\be\label{qcwf_f}
\psi (\varphi) = \left\{ \begin{array}{ll} A_0
\exp\bigg\{-\sqrt{\frac{E_J}{2E_C}} \bigg[\sqrt{1-\frac{\omega_p}{4E_J}}(\varphi-\pi)\bigg.\bigg.
, &  \varphi \lesssim \pi \\ \qquad \qquad
\bigg.\bigg. - \frac{(\varphi-\pi)^3}{24\sqrt{1-\omega_p/4E_J}} \bigg]\bigg\} & \\
A_0 \exp \left[ -\sqrt{\frac{E_J}{2E_C}}\sqrt{1-\frac{\omega_p}{4E_J}}
(\varphi - \pi )\right], & \varphi > \pi \end{array}\right.
\ee
with
\be
A_0 = \frac{1}{(2\pi)^{1/4}} \left(\frac{8E_J}{E_C}\right)^{1/8} e^{-\sqrt{2E_J/E_C}} \, .
\ee
The wavefunction near $\varphi = -\pi$ can be obtained by substituting $\varphi \to - \varphi$
in \eref{qcwf_f}.
We can now proceed with the calculation of the integral in \eref{te0}. Using \esref{atpot} and \rref{qcwf_f},
expanding the potential for $\varphi \le \pi$, and changing the integration
variable ($\varphi \to \pi - \varphi$) we find
\be\label{te0f}\begin{split}
\epsilon_0 & \simeq 2 E_J A_0^2 \int_0 \!d\varphi \, \varphi^2 \exp\left[-\sqrt{\frac{E_J}{2E_C}}\,
\frac{\varphi^3}{24\sqrt{1-\omega_p/4E_J}}\right]
 \\ & \simeq 8 \omega_p A_0^2
= 4\sqrt{\frac{2}{\pi}} \omega_p \left(\frac{8E_J}{E_C}\right)^{1/4} e^{-\sqrt{8E_J/E_C}}
\end{split}\ee
where, going from the first to the second line, we neglect the subleading correction originating from the
denominator in the argument of the exponential. The final expression for $\epsilon_0$ agrees
with the known asymptotic formula,\cite{transmon,flux_th,jcp} thus validating our approach.

The above result can be generalized to calculate the splitting between nearly degenerate even/odd states of
approximate energy $n \omega_p$ above the ground state
by letting $\omega_p \to \omega_p (2n+1)$ in \esref{qcwf}-\rref{p_def} and those that follow
[this replacement is appropriate so long as
$\omega_p(n+1/2) \ll 2E_J$]. Matching the semiclassical wavefunction to the excited eigenstates of the
harmonic oscillator, we find that
the normalization coefficient depends on $n$,
\be
C_n = \sqrt{\frac{\omega_p}{4E_C}}\left(\frac{2}{\pi e}\right)^{1/4}
\left(\frac{n+1/2}{e}\right)^{n/2} \left(\frac{\sqrt{n+1/2}}{n!}\right)^{1/2}.
\ee
Note that $C_n$ approaches $C_\infty$ as $n$ grows. Repeating the above calculation, we find
the energy splitting
\be
\epsilon_n = \epsilon_0 (-1)^n \frac{2^{2n}}{n!}
\left(\frac{8E_J}{E_C}\right)^{n/2}\, ,
\ee
also in agreement with the expression in the literature.

\section{Rate of parity switching induced by quasiparticles in the transmon}
\label{app:eorate}

The spectrum of the transmon, as described in Appendix~\ref{app:eosplit}, comprises both well separated
and nearly degenerate levels of opposite parity (see also Fig.~\ref{fig:trans}).
The leading contribution to the transition rate between
states of different parity separated in energy by (approximately) the plasma frequency
is given by \eref{Gnn} with $\varphi_0=0$ and is
independent of $n_g$. Here we consider the
quasiparticle-induced transitions between the nearly degenerate states $\Psi^e$ and $\Psi^o$.
We first consider a single-junction transmon to show explicitly that the rate depends on $n_g$ and
is exponentially small. Next we study the experimentally relevant case of a split transmon; its rate is
qualitatively different, not displaying such exponential smallness.

\subsection{Single-junction transmon}

According to \eref{wif_gen}, the quasiparticle transition rate $\Gamma_{o \to e}$ between states
$\Psi^o$ and $\Psi^e$ can be written as
\be\label{Geo}
\Gamma_{o \to e} = \left|\langle \Psi^e |\sin \frac{\hat\varphi}{2}| \Psi^o \rangle\right|^2
S_\qp\left(\omega_{eo}\right).
\ee
This rate depends on the gate voltage $n_g$ via the states in the matrix element as well as via their energy
difference $\omega_{eo}$, see \eref{de_ng}. For the matrix element
we use \eref{Psieo} to find
\be\label{eo_me_sjt}
\left|\langle \Psi^e |\sin \frac{\hat\varphi}{2}| \Psi^o \rangle\right| \simeq\left|\sin (2\pi n_g)\right| s \, ,
\ee
where
\be\label{sdef}
s = 2 \left| \int d\varphi \, \psi(\varphi) \psi(\varphi -2\pi) \sin \frac{\varphi}{2} \right| \, .
\ee
The matrix element in \eref{eo_me_sjt} vanishes at half integer values of $n_g$, as in the case of
the Cooper pair box [see \eref{cpb_me}]. In fact, the vanishing
holds at arbitrary ratio $E_J/E_C$, as can be shown using the
symmetry properties of Mathieu functions. For example, at $n_g=0$, $1/2$ the two lowest eigenstates of the
transmon Hamiltonian, \eref{Hphi_tr},
can be written in the charge basis as~\cite{mat_ord}
\be\begin{split}
|\Psi^e\rangle & = \sum_{m=0}^{\infty} A^{(0)}_{2m} \Big[|2m\rangle + |-2m\rangle \Big], \\
|\Psi^o\rangle & = \sum_{m=0}^{\infty} A^{(1)}_{2m+1} \Big[|2m+1\rangle + |-(2m+1)\rangle \Big],
\end{split}\ee
and
\be\label{eost_ngh}\begin{split}
|\Psi^e\rangle & = \sum_{m=0}^{\infty} A^{(1)}_{2m+1} \Big[|2m+2\rangle + |-2m\rangle \Big], \\
|\Psi^o\rangle & = \sum_{m=0}^{\infty} A^{(0)}_{2m} \Big[|2m+1\rangle + |-2m+1\rangle \Big],
\end{split}\ee
respectively, where the coefficients
$A^{(0)}_{2m}$, $A^{(1)}_{2m+1}$ depend on the ratio $E_J/E_C$.\cite{red_cpb}
Using the charge basis representation of $\sin\hat\varphi/2$ in \eref{sin_cb} it is easy to check
the vanishing of its matrix element between the above states for both values of $n_g$.

In the transmon limit $E_J/E_C\gg 1$ under consideration,
the product of wavefunctions localized at the same site does not contribute to the matrix
element in \eref{eo_me_sjt}: the intrawell integral vanishes because $\psi^2(\varphi)$ is a symmetric function
($\psi$ being the ground state of a symmetric potential) which is multiplied by the antisymmetric
function $\sin\varphi/2$; the vanishing of the intrawell term has thus he same origin of the vanishing of the
matrix element for a weakly anharmonic qubit at zero phase bias, see \eref{harm_mat_el} with $n=m$ and
$\varphi_0=0$.
To estimate the interwell contribution $s$ in \eref{sdef}, we use \eref{qcwf_f} and that
near $\varphi = \pi$ we have $\sin \varphi/2 \simeq 1$. After
changing integration variable ($\varphi \to \pi - \varphi$) we arrive at
\be\label{sfin}\begin{split}
s & \simeq 4 A_0^2 \int_0 \!d\varphi \, \exp\left[-\sqrt{\frac{E_J}{2E_C}}\,
\frac{\varphi^3}{24\sqrt{1-\omega_p/4E_J}}\right]\\ & \simeq
D \left(\frac{E_C}{E_J}\right)^{1/6} \frac{\epsilon_0}{\omega_p}\, ,
\end{split}\ee
where (with $\Gamma$ denoting here the gamma function)
\be\label{Ddef}
D = 2^{1/6} 3^{-2/3} \Gamma\left(\frac{1}{3}\right) \approx 1.45 \, .
\ee
Due to the factor $\epsilon_0$ in \eref{sfin}, the transition rate in \eref{Geo} is indeed
exponentially small. Turning now to the factor $S_\qp$ in \eref{Geo}, we note that its argument, $\omega_{eo}$,
is usually small due to its exponential suppression at large $E_J/E_C$, see \eref{te0f}.
Therefore the ``high frequency'' condition $\omega_{eo} \gg \delta E$ (with $\delta E$ the characteristic
quasiparticle energy) is in general not satisfied and use of \eref{Sqp_hf} expressing $S_\qp$
in terms of the quasiparticle density is not appropriate. In thermal equilibrium, one can use
\eref{Sqp_th} for arbitrary ratio $\omega_{eo}/T$. Assuming $\epsilon_0 \ll T$, using \eref{K0_as},
\eref{Sqp_th}, and the above results, we rewrite \eref{Geo} as
\be\label{Geo_th}\begin{split}
\Gamma_{o \to e} = &
\frac{16 E_J}{\pi} e^{-\Delta/T} \left[\ln\frac{4T}{|\epsilon_0\cos(2\pi n_g)|} - \gamma_E \right]
\\ & \times
\left(\frac{E_C}{E_J}\right)^{1/3}\left(D\frac{\epsilon_0}{\omega_p}\right)^2
\sin^2 \left(2\pi n_g\right) \, .
\end{split}\ee
Generalization of this result to the transition rate $\Gamma^{(n)}_{o\to e}$ between nearly degenerate states
of higher
energy is obtained by the substitution $\epsilon_0\to \epsilon_n$.
Except at the degeneracy points $n_g=1/4$, $3/4$ (where this expression diverges), we can estimate
the rate in order of magnitude by assuming $\sin(2\pi n_g)$, $\cos(2\pi n_g) \approx 1$. For low-lying
states, this estimate shows that the rate $\Gamma^{(n)}_{o \to e}$ is small compared to the rate
$\Gamma_{1\to 0}$ determining the relaxation time of the transmon [see \eref{Gnn}]. This smallness
is due to the exponentially suppressed $o\to e$ matrix element, \eref{sfin}, as function of the
ratio $E_J/E_C$, in comparison with the weak power-law suppression of the $1\to 0$ matrix element
as given by \eref{harm_mat_el} with $\varphi_0=0$, $m=1$, and $n=0$.
The relationship between the two rates is qualitatively different in the split
transmon, as we discuss next.

\subsection{Split transmon}
\label{app:eorate-sp}

The above calculation of the even/odd transition rate in the single junction transmon
can be easily modified to yield the rate for a split transmon. As discussed in Sec.~\ref{sec:split_tr},
the effective Hamiltonian and therefore the form of the eigenstates are the same in the single and split
transmon. The difference between the two cases arises in the evaluation of the matrix elements pertaining
to each junction [cf. \eref{sp_tr_me0}]. For the even/odd matrix element we find
\be\label{eo_me_sp}
\left|\langle \Psi^e |\sin \frac{\hat\varphi_{j}}{2}| \Psi^o \rangle\right|^2 \simeq
\frac{1-\cos(\pi f\pm \vartheta)}{2} \, ,
\ee
where the upper (lower) sign applies to junction $j=1$ ($j=0$), $f$ is defined in \eref{min_pos}
and $\vartheta$ in \eref{vt_def}. In contrast with the single-junction transmon case considered above,
here the matrix element is dominated by the intrawell contribution having the same form
of \eref{harm_mat_el} at $n=m=0$ and finite phase bias $\pi f \pm \vartheta$.
Substituting \eref{eo_me_sp} into \eref{wif_multi} and assuming
thermal equilibrium quasiparticles [cf. \eref{Sqp_th}] we obtain
\be\label{Geo_sp}\begin{split}
\Gamma_{o \to e} = \frac{8 (E_{J0}+E_{J1})}{\pi}
e^{-\Delta/T} e^{\omega_{eo}/2T}K_0\left(\frac{|\omega_{eo}|}{2T}\right) \\
\left(1- \frac{\omega_p^2(f)}{\omega_p^2(0)}\right)
\end{split}\ee
with $\omega_p(f)$ given in \eref{opf_def} and $\omega_{eo}$ in \eref{de_ng}.
As before, the rate $\Gamma^{(n)}_{o \to e}$ of transitions between nearly degenerate levels of higher energy
is obtained upon the substitution $\omega_{eo} \to \epsilon_n \sin(2\pi n_g)$ in \eref{Geo_sp}.
Note that the rate vanishes at integer multiples of
the flux quantum; at those values of flux, exponentially small contributions to the
matrix element analogous to those calculated above should be included. At non-integer
values of reduced flux $f$,
\eref{Geo_sp} should be compared with the transition rate between qubit states induced by thermal
quasiparticles,
\bea
\Gamma_{1 \to 0} & = & \frac{8 (E_{J0}+E_{J1})}{\pi}
e^{-\Delta/T} e^{\omega_p(f)/2T}K_0\left(\frac{|\omega_p(f)|}{2T}\right) \nonumber \\
& & \frac{E_C}{\omega_p(f)}\left(1 + \frac{\omega_p^2(f)}{\omega_p^2(0)}\right),
\eea
obtained using \eref{sp_tr_me0}. The ratio between these two quantities,
\be\label{Grat}
\frac{\Gamma_{o \to e}}{\Gamma_{1 \to 0}} = \frac{e^{\omega_{eo}/2T}K_0\left(\frac{|\omega_{eo}|}{2T}\right)}
{e^{\omega_p(f)/2T}K_0\left(\frac{|\omega_p(f)|}{2T}\right)}\frac{\omega_p(f)}{E_C}
\frac{\omega_p^2(0) - \omega_p^2(f)}{\omega_p^2(0) + \omega_p^2(f)}
\ee
depends on temperature through the first factor on the right hand side. Experimentally, measurements for the
rate are performed near $n_g =1/2$, so that the relevant even/odd frequencies are $\omega_{eo} \sim
\epsilon_0, ~\epsilon_1$; they are generally 2-3 orders of magnitude smaller than $\omega_{p}(f)$
($\sim 2\pi \times 4$~GHz),
while the latter is usually larger than twice the temperature ($T\sim 20-200$~mK).
Under these conditions, the
first factor in \eref{Grat} can be approximated, in order of magnitude, by 5 to 10. The last factor in
\eref{Grat} varies between $0$ at $f=0$ and $1$ at $f=1/2$; as flux is used to suppress
the qubit frequency from its maximum value ($\gtrsim 10$~GHz), we can approximate the last factor by 1/2. Finally,
the central factor can be rewritten as $\sqrt{8E_J(f)/E_C}$; since $E_J(f)/E_C$ usually is varied between
10 and 30, we arrive at the order-of-magnitude estimate
\be
\frac{\Gamma_{o \to e}}{\Gamma_{1 \to 0}} \sim 20-80
\ee
in the experimentally relevant ranges of parameters. This is an example of the more general statement that,
except close to integer values of $f$, the even/odd transition rate in a split transmon is faster than
its decay rate.
This result is qualitatively in agreement with experimental bounds for the even/odd
transition rate in split transmons.~\cite{schreier,sun}

\section{Matrix elements for the harmonic oscillator}
\label{app:home}

In this Appendix we present analytic expression for the matrix elements of $\sin\hat\varphi/2$ between
harmonic oscillator states $|n\rangle$ and $|m\rangle$. Let us introduce the displacement operator
\be
\hat{D}(\mu) = e^{\mu \hat a^\dagger - \mu^* \hat a} \, ,
\ee
where $\hat a$ ($\hat a^\dagger$) is the harmonic oscillator annihilation (creation) operator.
The matrix elements of $\hat D$ are~\cite{Laguerre}
\be\label{meD}
\langle m | \hat D(\mu) |n\rangle =
\left\{\begin{array}{r}
e^{-|\mu|^2/2} \sqrt{\frac{m!}{n!}} (-\mu^*)^{n-m} L^{(n-m)}_m \left(|\mu|^2\right), \\
m\leq n \\
e^{-|\mu|^2/2} \sqrt{\frac{n!}{m!}} (\mu)^{m-n} L^{(m-n)}_n \left(|\mu|^2\right),
\phantom{-}\\
m \geq n
\end{array}\right.
\ee
where $L^{(\alpha)}_n$ are the generalized Laguerre polynomials.
Since the position operator is $\hat\varphi = \ell (\hat a + \hat a^\dagger)/\sqrt{2}$, where
$\ell = 1/\sqrt{m\omega}$ is the oscillator length for an oscillator of mass $m$ and frequency $\omega$,
we can write
\be\label{eD}
e^{i\hat\varphi/2} = \hat D\left(\frac{i \ell}{2\sqrt{2}}\right)\, .
\ee
Note that for the harmonic oscillator described by \eref{Hphi_2} we have
\be
\ell = 2\sqrt{2} \sqrt{\frac{E_C}{\EZO}} \, .
\ee

To allow for fluctuations around a finite phase, we shift $\hat\varphi \to \varphi_0 + \hat\varphi$
in the argument of sine and rewrite the resulting expression in terms of exponentials,
\be
\sin \frac{\varphi_0 + \hat\varphi}{2} = \frac{1}{2i}\left(
e^{i\varphi_0/2} e^{i\hat\varphi/2} - e^{-i\varphi_0/2} e^{-i\hat\varphi/2}\right) \, .
\ee
Then using \esref{meD}-\rref{eD} we find for $m\leq n$
\be\label{full_home}\begin{split}
\langle m | \sin \frac{\varphi_0 + \hat\varphi}{2} |n\rangle =
e^{-\ell^2/16} \sqrt{\frac{m!}{n!}}\left(\frac{\ell}{2\sqrt{2}}\right)^{n-m}
\\ \times
L_m^{(n-m)}\left(\frac{\ell^2}{8}\right)
\sin\frac{\varphi_0 +\pi(n-m)}{2}\, .
\end{split}\ee
The matrix element for $m\geq n$ is obtained by exchanging $n \leftrightarrow m$ in the right hand side.
Equation \rref{harm_mat_el} can be obtained from \eref{full_home} by Taylor expansion for small
$\ell$, which for the Laguerre polynomials gives
\be
L_m^{(\alpha)}(x) = \frac{(m+\alpha)!}{m!\alpha!} -\frac{(m+\alpha)!}{(m-1)!(\alpha+1)!} x + {\cal O}(x^2) .
\ee
Equation \rref{me_zn} follows from \eref{full_home} with $m=0$
using $L_0^{(\alpha)}(x) = 1$.

Using \eref{meD} we can also find the expectation value of the operator $\cos \hat\varphi$. After shifting the
phase variable as done above and since the expectation value of sine vanishes by symmetry, we find
\be
\langle n | \cos \left(\varphi_0 + \hat\varphi\right) |n\rangle = \cos\varphi_0
\langle n | \cos \hat\varphi |n\rangle \, .
\ee
Writing the cosine in exponential form, using $e^{i\hat\varphi} = \hat D\left(i \ell/\sqrt{2}\right)$
we arrive at
\be\label{me_cos}
\langle n | \cos \left(\varphi_0 + \hat\varphi\right) |n\rangle = \cos\varphi_0 \,
e^{-\ell^2/4} L_n^{(0)} \left(\frac{\ell^2}{2}\right) \, .
\ee

\section{Matrix elements for the transmon}
\label{app:tme}

Here we want to show that corrections to \eref{harm_mat_el} for the transmon ($\varphi_0 =0$) are of
cubic order in $E_C/\omega_p$, as claimed in the text following that equation. The transmon
Hamiltonian is given by
\eref{Hphi_tr} and we neglect exponentially small corrections by setting $n_g =0$ (see
\ocite{transmon} and Appendices~\ref{app:eosplit} and \ref{app:eorate}).
Numbering the eigenstates $|\psi_n\rangle$ starting with $n=0$ for the ground state,
even (odd) numbered states are even (odd) functions of $\varphi$, due to the symmetry of the potential energy.
Since $\sin\varphi/2$ is an odd function,
the matrix element between states of the same parity vanishes,
\be
\langle \psi_{n\pm 2j} | \sin \frac{\hat\varphi}{2} | \psi_n \rangle = 0 \, , \qquad j=0,1,2,\ldots
\ee

Due to the smallness of the charging energy, $E_C \ll E_J$, as a first approximation
we can expand the Josephson energy in \eref{Hphi_tr} up to the fourth order in $\varphi$.
In terms of creation/annihilation operators (cf. Appendix~\ref{app:home} -- note that in the
present case $\ell = 2\sqrt{2} \sqrt{E_C/\omega_p} \ll 1$), the approximate transmon Hamiltonian is
\bea
\hat{H} & = & \hat H_0 + \delta \hat H \, , \\
\hat H_0 & = & \omega_p \left(\hat a^\dagger \hat a + \frac12\right) \, , \\
\delta \hat H & = &
- \frac{E_C}{12}\left(a+a^\dagger\right)^4 \, . \label{dH}
\eea
To first order in $E_C/\omega_p$, expressed in terms of harmonic oscillator states the transmon eigenstates are
therefore
\bea\label{psiapp}
|\psi_n \rangle &=& |n\rangle + |\delta \psi_n \rangle\,, \\
|\delta \psi_n \rangle &=&  - \sum_{j\neq n} |j\rangle
\frac{\langle j | \delta \hat H | n \rangle}{E_j - E_n} \, , \qquad E_n = \omega_p\left(n +\frac12\right),
\nonumber
\eea
and including the first anharmonic corrections to the eigenstates the matrix elements are
\be\label{tr_me_1}\begin{split}
\langle \psi_{m} | \sin \frac{\hat\varphi}{2} | \psi_n \rangle \simeq
\langle m| \sin \frac{\hat\varphi}{2} | n \rangle  +
\langle m | \sin \frac{\hat\varphi}{2} | \delta \psi_n \rangle \\ +
\langle \delta \psi_{m} | \sin \frac{\hat\varphi}{2} | n \rangle \, .
\end{split}\ee
Using \eref{full_home}, we find that the leading contribution to the first term on the right hand side is
\be\label{tr_me_j}
\langle n \pm (2j+1) | \sin \frac{\hat\varphi}{2} | n \rangle \propto
\left(\frac{E_C}{\omega_p}\right)^{j+1/2} , \quad j=0,1,2,\ldots
\ee
Since we are interested in calculating the
square of the matrix elements up to second order in $E_C/\omega_p$,
we can neglect transitions with $j \geq 1$. For $j=0$, using \eref{full_home} at next to leading order we find
\be\label{home}\begin{split}
& \langle n \pm 1 | \sin \frac{\hat\varphi}{2} | n \rangle \simeq
\sqrt{\left(n+\frac12 \pm \frac12 \right) \frac{E_C}{\omega_p}}
\\ & \, \times \left[1 - \frac12
\left(n+\frac12 \pm \frac12 \right) \frac{E_C}{\omega_p}+{\cal O}\left(\frac{E_C}{\omega_p}\right)^2\right].
\end{split}\ee

Consider now the case $m=n-1$ in \eref{tr_me_1}. Using \esref{dH}, \rref{psiapp},
and the leading term in \eref{home},
the central term in the right hand side is approximately
\be\label{cme1}\begin{split}
\langle n-1 | \sin \frac{\hat\varphi}{2} | \delta \psi_n \rangle
\simeq -\langle n-1 | \sin \frac{\varphi}{2} | n-2 \rangle
\frac{\langle n-2 | \delta \hat H | n \rangle}{E_{n-2} - E_n} \\
\simeq -\frac{1}{24}\left(\frac{E_C}{\omega_p}\right)^{3/2}\sqrt{n-1}
\langle n-2 | \left(a+a^\dagger\right)^4 | n \rangle \, .
\end{split}\ee
To calculate the last factor we note that
\be\begin{split}
\left(a+a^\dagger\right)^2 | n \rangle =
\sqrt{n(n-1)}|n-2\rangle + (2n+1)|n\rangle \\ + \sqrt{(n+1)(n+2)}|n+2\rangle .
\end{split}\ee
Shifting $n\to n-2$ and taking the scalar product we arrive at
\be
\langle n-2 | \left(a+a^\dagger\right)^4 | n \rangle = 4 \sqrt{n(n-1)} \left(n-\frac12\right),
\ee
and substituting this expression into \eref{cme1} we obtain
\be\label{dpn}
\langle n-1 | \sin \frac{\hat\varphi}{2} | \delta \psi_n \rangle =
-\frac16 \left(\frac{E_C}{\omega_p}\right)^{3/2} \sqrt{n} (n-1) \left(n-\frac12\right).
\ee
Proceeding as above we also find
\be\label{dpnm}
\langle \delta\psi_{n-1} | \sin \frac{\hat\varphi}{2} | n \rangle =
\frac16 \left(\frac{E_C}{\omega_p}\right)^{3/2}
\sqrt{n} (n+1) \left(n+\frac12\right).
\ee
Finally, substitution of \esref{home}, \rref{dpn}, and \rref{dpnm} into \eref{tr_me_1} gives
\be
\langle \psi_{n-1} | \sin \frac{\hat\varphi}{2} | \psi_n \rangle =
\sqrt{n \frac{E_C}{\omega_p}}
+ {\cal O}\left(\frac{E_C}{\omega_p}\right)^{5/2}.
\ee

Repeating the above calculations for the case $m=n+1$ and using \eref{tr_me_j} we conclude that
the square of the matrix element is
\bea
\left|\langle \psi_{m} | \sin \frac{\hat\varphi}{2} | \psi_n \rangle\right|^2 & = & \frac{E_C}{\omega_p}
\left[n\delta_{m,n-1} + (n+1)
\delta_{m,n+1}\right]  \nonumber \\ & & + {\cal O}\left(\frac{E_C}{\omega_p}\right)^{3}.
\eea

\section{Multi-junction Hamiltonian}
\label{app:mj_ham}

The aim of this Appendix is to
derive the Hamiltonian for a multi-junction system starting from the Lagrangian, \eref{lagra}.
We consider a loop of $M+1$ junctions and assume $M$ of them, denoted
by index $j$ with $j=1,\ldots,M$, to be identical, so that their capacitances and Josephson energies
are, respectively, $C_j = C_1$ and $E_{Jj} = E_{J1}$ for $1\leq j \leq M$. These $M$ junctions
will be referred to as the array junctions to distinguish them
from the $j=0$ junction, whose capacitance $C_0$ and Josephson energy $E_{J0}$ can differ from those
of the array junctions.

While the system comprises $M+1$ junctions, there are only $M$ independent degrees of freedom, due to
the flux quantization constraint, \eref{flux_quant}. Using that equation to eliminate the phase $\varphi_0$,
the Lagrangian is
\bea
&& {\cal L}_{\{\varphi\}} =\frac{1}{2} \frac{C_0}{(2e)^2}\left(\sum_{j=1}^M\dot{\varphi}_j\right)^2 +
\frac{1}{2} \frac{C_1}{(2e)^2} \sum_{j=1}^M \dot{\varphi}_j^2
\\ && + E_{J0} \cos \left(\sum_{j=1}^M \varphi_j - 2\pi \Phi_e/\Phi_0 \right)
+E_{J1} \sum_{j=1}^M \cos \varphi_j \, . \nonumber
\eea
We introduce a new set $\{\phi\}$ of $M$ independent variables
\bea\label{nv}
\phi_{\phantom{k}} & = & \sum_{j=1}^M \varphi_j \, , \\
\phi_k & = & \varphi_k - \alpha\sum_{l=1}^{M-1} \varphi_l
+\frac{\varphi_M}{\sqrt{M}} \, , \quad k=1,\ldots,M-1 \, ,\qquad
\eea
where
\be\label{al_def}
\alpha = \left(1+\frac{1}{\sqrt{M}}\right)\frac{1}{M-1}.
\ee
The inverse transformation is given by
\bea
\varphi_k & = & \phi_k - \alpha \sum_{l=1}^{M-1} \phi_l
+\frac{1}{M}\phi \, , \qquad k=1,\ldots,M-1,  \nonumber \\
\varphi_M & = & \frac{1}{\sqrt{M}} \sum_{l=1}^{M-1}\phi_l + \frac{1}{M}\phi \, . \label{nvi}
\eea
In terms of the $M$ variables $\phi$, $\phi_k$ ($k=1,\ldots, M-1$) the Lagrangian is
\be
{\cal L}_{\{\phi\}} = \frac{1}{8e^2}\left(C_0 + \frac{C_1}{M}\right)\dot{\phi}^2 +
\frac{1}{8e^2} C_1 \sum_{k=1}^{M-1}\dot\phi_k^2 - U(\{\phi\})
\ee
with potential energy
\be\label{Udef}\begin{split}
U(\{\phi\}) =&
-E_{J0} \cos\left(\phi - 2\pi\Phi_e/\Phi_0\right)
\\ & -  E_{J1} \sum_{k=1}^{M-1} \cos\left(\phi_k -\alpha\sum_{l=1}^{M-1}\phi_l + \frac{\phi}{M}\right)
\\ & - E_{J1} \cos \left(\frac{1}{\sqrt{M}}\sum_{l=1}^{M-1} \phi_l + \frac{\phi}{M}\right).
\end{split}\ee

Introducing the $M$ conjugate variables $N = \partial {\cal L}_\phi/\partial \phi$ and
$N_k = \partial {\cal L}_\phi/\partial \phi_k$ ($k=1,\ldots,M-1$), the Hamiltonian is
\be\label{Hm_ex}\begin{split}
H_{\{\phi\}} & = N\dot\phi + \sum_{k=1}^{M-1} N_k\dot\phi_k - {\cal L}_{\{\phi\}} \\
& = 4E_C N^2 + 4E_{C1} \sum_{k=1}^{M-1}N_k^2 + U(\{\phi\}),
\end{split}\ee
where
\be\label{mj_ec}
E_{C}  = \frac{e^2}{2(C_0 + C_1/M)}\, , \qquad
E_{C1}  = \frac{e^2}{2C_1}\, .
\ee
The Hamiltonian in \eref{Hm_ex} governs the dynamics of the $M$ independent degrees of freedom
of the $M+1$ junction system with flux quantization and $M$ identical array junctions.
For a two junction system we have $M=1$ and all the sums in \esref{Udef}-\rref{Hm_ex} are absent. Then the
Hamiltonian is that given in \eref{tr_sp_ham0}.

\subsection{Fluxonium}

The fluxonium consist of $M+1$ junctions such that a ``weak'' junction $j=0$ with $E_{J0} < E_{J1}$
is connected to a large
array of $M$ junctions ($M \gg 1$) with small phase fluctuations, $E_{C1} \ll E_{J1}$.
These conditions enable us to drastically simplify
the last two terms of the potential energy $U(\{\phi\})$ for the
$M$ independent variables $\phi$, $\phi_k$ ($k=1,\ldots, M-1$) in \eref{Udef}.

We consider small fluctuations of variables $\phi_k$
around the configuration $\phi_k =0$, $k=1,\ldots,M-1$, which is an extremum of $U$ for any value of $\phi$
[as can be checked by differentiating $U$ with respect to $\phi_k$ and using \eref{al_def}].
We further assume that typical values of $\phi$ are small compared to $2\pi M$ (note that
since $M$ is large, this weak restriction on $\phi$ and its fluctuations
still allows for phase slips through the weak junction). Then we can expand the
last two terms in \eref{Udef} to quadratic order in $\phi_k$ and $\phi/M$ to find
\be\begin{split}
U(\{\phi\}) \simeq & -E_{J0} \cos\left(\phi - 2\pi\Phi_e/\Phi_0\right) + \frac12 E_L \phi^2
\\ & +\frac 12 E_{J1} \sum_{k=1}^{M-1} \phi_k^2
\end{split}\ee
with
\be\label{mj_el}
E_L =\frac{E_{J1}}{M} \, .
\ee
Hence in this approximation the Hamiltonian \rref{Hm_ex} for the $M+1$ junction fluxonium
separates into independent Hamiltonians for
each of the $M$ unconstrained variables $\phi$, $\phi_k$,
\bea
H_{\{\phi\}} & = &  H_\phi + \sum_{k=1}^{M-1} H_k \, ,\\
H_\phi & = & 4E_CN^2 -E_{J0}\cos(\phi-2\pi\Phi_e/\Phi_0) + \frac12 E_L\phi^2 , \nonumber \\
H_k &=& 4E_{C1}N_k^2 + \frac12 E_{J1} \phi_k^2 \, .\nonumber
\eea
Up to a change of variable $\phi \to 2\pi\Phi_e/\Phi_0 - \phi$
and redefinitions of symbols, $H_\phi$ coincides with $H_\varphi$ of \eref{Hphi}. The relations
in \eref{fl_par} between the parameters of the $M+1$ junctions and the energies $E_C$ and $E_L$
entering the effective qubit Hamiltonian $H_\phi$ follow from
\esref{mj_ec} and \rref{mj_el}, respectively.


\begin{thebibliography}{50}
\bibitem{noteEL}More precisely, we need to assume that low-lying states in a well are separated
from those in nearby wells. This means that we neglect tunneling between wells [which is suppressed when
$E_J \gg E_C$, see \eref{ep0_def}].
Tunneling is taken into account in Sec.~\ref{sec:fl_qb}.
\bibitem{note1}Here and below, in writing the conditions for the validity of our approximations we neglect
factors which, for the flux-biased phase qubit, depends on $\varphi_0$. This dependence is through
trigonometric functions with $\varphi_0$ as argument, and for typical flux biases these functions
would modify the applicability conditions by
factors of order unity. This is consistent with our assumption that the flux is sufficiently far from
odd integer multiples of half the flux quantum.
\bibitem{k_corr}In the case of the transmon, an estimate for the rate $\Gamma_{1\to 0}$
was given in Eq.~(4.10) of
\ocite{transmon} by generalizing the results obtained in \ocite{lutchyn2} for the Cooper pair box.
Note that the present more rigorous treatment gives a parametrically larger rate (by a factor
$\sim \sqrt{\Delta/T}$) when applying \eref{Gnn} to a single junction transmon.
\bibitem{cpbcomp}We note that in \ocite{flux_th} the limit $E_L \to 0$ was considered at a fixed
ratio $E_J/E_C$.
\bibitem{mf}The factors in the second line in \eref{G_fl_q} were missed in Eq.~(22) of \ocite{prl}. Those
factors increase the magnitude of the transition rate but do not affect its functional dependence on flux $f$.
\bibitem{ngnote}In this section we neglect for simplicity the dimensionless gate voltage $n_g$.
As discussed in Sec.~\ref{sec:semi}, this amounts to neglecting exponentially small corrections,
since for both split transmon and fluxonium we consider the regime of Josephson energy large
compared to charging energy.
\bibitem{ren_neg}Here we neglect for simplicity the renormalization of the parameters that appear in
\eref{fl_qb_me}. This is justified since its contribution to the rate can be neglected -- see the main
text.
\bibitem{renorm}In the terms we are neglecting we can distinguish contributions proportional to the
distribution function $f$ from a contribution independependent of $f$. The contribution
proportional to $f$ originating from $\delta E_i^{(1),d}$ is logarithmically
divergent, but this divergence is cancelled by the similar contribution from $\delta E_i^{(2)}$.
The contribution independent of $f$ in $\delta E_i^{(2)}$ is linearly divergent, but independent of
the gap $\Delta$; threfore, this divergence can be eliminated by considering energy differences between
the superconducting and normal state.
\bibitem{red_cpb}Note that the two lowest eigentstates
of the reduced Cooper pair box Hamiltonian \rref{HCPB} at $n_g=1/2$ -- see \esref{st_o} and \rref{st_e} --
have the form of the $m=0$ contributions to the sums in \eref{eost_ngh}.
\end{thebibliography}
\end{document}